\documentclass[12pt]{article}

        \usepackage{amsfonts}
        \usepackage{amssymb}
    \usepackage{graphics}

\input epsf.def

\newtheorem{thm}{Theorem}
\newtheorem{cor}[thm]{Corollary}
\newtheorem{lemma}[thm]{Lemma}

\def\pf{\medbreak\noindent{\bf Proof:}\enspace}
\def\qed{~~{\bf QED}}
\def\be{\begin{eqnarray}}
\def\ee{\end{eqnarray}}
\def\bee{\begin{eqnarray*}}
\def\eee{\end{eqnarray*}}

        \def\iff{\Leftrightarrow}
   \def\imp{\Rightarrow}
        \def\tr{\hbox{Tr}}
        
      \def\Hil{{\cal H}}

\def\bu{{\bf u}}

\def\bw{ {\bf w}}
\def\bx{ {\bf x}}
\def\bu{ {\bf u}}

\def\bt{ {\bf t}}
\def\ot{\otimes}

\def\holv{{\rm Holv}}

\def\ds{\displaystyle}
 
\def\bra{\langle}
\def\ket{\rangle}
\def\kb{ \ket \bra }

\def\rt2{ \frac{1}{\sqrt{2}} }

\def\raw{\rightarrow}

\def\wh{\widehat}
\def\dg{\dagger}
\def\dtsig{{\mathbf \cdot \sigma}}
\def\rmT{{\rm T}}

 \def\cb{{\cal B}}
 \def\ca{{\cal A}}

\def\cs{{\mathcal{S}}}
\def\cd{{\mathcal{D}}}
\def\calm{{\mathcal{M}}}

\def\half{{\textstyle \frac{1}{2}}}

\def\thrd{{\textstyle \frac{1}{3}}}

     \title{An Analysis of Completely-Positive Trace-Preserving
Maps on $\mathcal{M}_2$}

        \author {Mary Beth Ruskai \thanks{Partially supported  by
 the National Security Agency and
 Advanced Research and Development Activity under
Army Research Office  contract 
   {DAAG55-98-1-0374} and by the National Science
        Foundation under grants DMS-9706981 and DMS-0074566.}
 \\ Department of Mathematics \\ University of Massachusetts  Lowell \\
Lowell, MA  01854 USA \\ {\normalsize bruskai@cs.uml.edu} \and
 Stanislaw Szarek \thanks{Partially supported  by
a grant from the National Science
        Foundation.}
\\Department of Mathematics \\
    Case Western Reserve University \\
    Cleveland, OH 44106 USA \\
    {\normalsize sjs13@po.cwru.edu}\\
     and \\ Equipe d'Analyse, Boite 186,
     Universit\'{e} Paris VI \\ 4, Place Jussieu,  F-75252  Paris, France \\
   {\normalsize szarek@ccr.jussieu.fr}
\and Elisabeth Werner
\thanks{Partially supported  by
a grant from the National Science
         Foundation and by a NATO Collaborative Linkage Grant}
\\Department of Mathematics \\
    Case Western Reserve University \\
    Cleveland, OH 44106 USA \\
    {\normalsize emw2@po.cwru.edu}\\
    and\\
    Universit\'{e} de Lille 1,
    UFR de Math\'{e}matique\\
    F-59655 Villeneuve d'Ascq, France}

\date{7 August 2001}

\begin{document}

 \maketitle

\begin{abstract}
We give a useful new characterization of the set of all completely
positive, trace-preserving maps
$\Phi: \mathcal{M}_2 \rightarrow \mathcal{M}_2$ from which
one can easily check any trace-preserving map for complete
positivity.
We also determine explicitly all extreme points of this set,
and give a useful parameterization after reduction to a
certain canonical form.  This allows a detailed examination
of an important class of non-unital extreme points that can be
characterized as having exactly two images on the Bloch sphere.

We also discuss a number of related issues about the images
and the geometry of the set of stochastic maps, and show that
any stochastic map on $\calm_2$ can be written as a convex combination
of two ``generalized" extreme points.
\end{abstract}

\small{\noindent {\bf Key Words:}  Completely positive maps, stochastic
maps,  
 quantum communication, noisy channels, Bloch sphere, states, .

\smallskip

\noindent{\bf MR Classification.}  47L07, 81P68,  
 15A99, 46L30, 46L60, 94A40 }

\tableofcontents

 \pagebreak

\bigskip

\section{Introduction}

\subsection{Background}

Completely positive, trace-preserving maps arise naturally in
quantum information theory and other situations in which one
wishes to  restrict attention to a quantum system that should
properly be considered a subsystem of a larger system with which
it interacts.  In such situations, the system of interest is
described by a Hilbert space $\Hil_1$ and the larger system by
a tensor product $\Hil = \Hil_1 \otimes \Hil_2$.
States correspond to density matrices, where a density matrix
$\rho$ is a positive semi-definite  operator on $\Hil_1$ with
$\tr \rho = 1$.  The result of the ``noisy'' interaction with
the larger system is described by a {\em stochastic} map
$\Phi: \cb(\Hil_1) \raw \cb(\Hil_1)$ that takes $\rho$ to another
density matrix $\Phi(\rho)$.  Since  $\Phi(\rho)$ should also be a
density matrix, $\Phi$ must be both trace-preserving and
positivity preserving.  However, the latter is not enough,
since $\Phi(\rho)$ is the result of a positivity-preserving
process on the larger space of operators in
$\cb(\Hil_1 \otimes \Hil_2) = \cb(\Hil_1) \otimes \cb(\Hil_2)$.
This is precisely the condition that $\Phi$ be completely positive.

The notion of complete positivity was introduced in the more general
context of linear maps $\Phi: \ca_1 \raw \ca_2$ on $C^*$-algebras
where the condition can be stated as the requirement that
$\Phi \otimes I_{\mathcal{M}_{n}} : \ca_1 \otimes \mathcal{M}_n \raw \ca_2
\otimes \mathcal{M}_n$
is positivity preserving for all $n$, where
$\mathcal{M}_n$ denotes the algebra of complex $n \times n$ matrices.
Stinespring showed that a completely positive map always has a
representation of the form $\pi_2[\Phi(A)] = V^* \pi_1(A) V$ where
$\pi_1$ and
$\pi_2$ are representations of the algebras $\ca_1, \ca_2$
respectively.  Kraus \cite{K70,Kbk} and Choi \cite{Choi1} showed
that this leads to the more tangible condition that there exists
a set of operators $A_k$ such that
\be \label{eq:kraus}
\Phi(\rho) = \sum_k A_{k}^{\dg} \rho A_{k}.
\ee
(where we henceforth follow the physics convention of using
$\dg$ to denote the adjoint of an operator.)
The condition that $\Phi$ is trace-preserving can then be
written as $\sum_k A_{k}  A_{k}^{\dg} = I$.
When this condition is also satisfied, (\ref{eq:kraus}) can also
be used \cite{Kbk,Lind} to find a representation of $\Phi$ in terms of
a partial trace  on a larger space.

The operators in (\ref{eq:kraus}) are often referred to as
``Kraus operators''  because of his influential work
\cite{K70,Kbk} on quantum measurement in which he emphasized the
role of completely positive maps.
 Recognition that such maps play a natural role
in the description of quantum systems goes back at least
to Haag and Kastler \cite{HK}.
 It is worth noting that this
representation is highly non-unique, and that this
non-uniqueness can not be eliminated by simple constraints.
In particular,  one can find extreme maps with at least  two
two different Kraus representations each of which uses only
the minimal number of Kraus operators.  An example is given
in Section \ref{sect:capac}.
The representation (\ref{eq:kraus}) was obtained independently
by Choi \cite{Choi1} in connection with  important tests,
upon which this paper is based, for
complete-positivity and extremality in the case of maps
on $\mathcal{M}_n$.

However, Choi's condition and all of the representations
discussed above require, at least implicitly, consideration of the map
$\Phi \otimes I_{\mathcal{M}_n}$ on a larger space.  (In quantum
information theory this problem is sometimes avoided by defining a
stochastic map, or channel, in terms of its Kraus operators as in
(\ref{eq:kraus}); however, this approach has thus tended to
focus attention on  a rather
limited set of channels.)   One would like to find
 a simple way to characterize completely positive maps in
terms of their action on the algebra $\cb(\Hil_1)$ of the
subsystem.  The purpose of this paper is to  obtain such a
characterization in the special case of trace-preserving maps
on $\cb(\Hil_1) = \mathcal{M}_2$.  This  leads to a complete
description of their extreme points, and a useful parameterization
of the stochastic maps and their extreme points.
Although the two-dimensional case $\Hil_1 = {\bf C}^2$ may
seem rather special, it is of considerable importance because
of its role in quantum computation and quantum communication.

If, in addition to being trace-preserving, a completely positive
map $\Phi$ is unital, i.e., $\Phi(I) = I$, we call $\Phi$
{\em bistochastic}.  This terminology for maps that are
both unital and stochastic was introduced in \cite{AHW}.

\subsection{Notation} \label{sect:not}

First, we note that for a linear map $\Phi$, its adjoint,
which we denote $\wh{\Phi}$ (to avoid confusion with the
operator adjoint of a specific image) can be defined
with respect to the Hilbert-Schmidt inner product
$\bra A, B \ket = \tr A^{\dg} B$ so that
$\tr \, [\Phi(A)]^{\dg} B = \tr \, A^{\dg} \wh{\Phi}(B)$.
It is easy to verify that the Kraus operators for
$\wh{\Phi}$ are the adjoints of those for $\Phi$ so that
(\ref{eq:kraus}) implies
$\wh{\Phi}(\rho) = \sum_k A_{k} \rho A_{k}^{\dg}$, and that
$\Phi$ is trace-preserving if and only if
$\wh{\Phi}(I) = I$,  i.e., if $\wh{\Phi}$ is unital.

In order to state our results in a useful form, we recall
 that the identity and Pauli matrices
$\{ I, \sigma_x, \sigma_y, \sigma_z \}$ form a basis
for $\mathcal{M}_2$ where
\begin{eqnarray*}
\sigma_x = \left( \begin{array}{cc} 0 & 1 \\ 1 & 0 \end{array} \right)~~
\sigma_y = \left( \begin{array}{cc} 0 & -i \\ i & 0 \end{array} \right)~~
\sigma_z = \left( \begin{array}{cc} 1 & 0 \\ 0 & -1 \end{array} \right).
\end{eqnarray*}
Every density matrix can  be written
 in this basis
as $\rho = {\half} [I + \bw \cdot {\bf \sigma}]$ with
$\bw \in {\bf R}^3$ and $|\bw| \leq 1$.   Thus, the set of
density matrices, which we shall denote by $\cd$,
can be identified with the unit ball in
${\bf R}^3$  and the pure states (rank one projections)
lie on the surface known as the ``Bloch sphere.''   Since
$\Phi$ is a linear map on $\mathcal{M}_2$, it can also be represented
in this basis by a unique $4 \times 4$
matrix ${\bf T}$, and $\Phi$ is trace-preserving if
and only if the first row satisfies $t_{1k} = \delta_{1k}$, i.e.,
${\bf T} =  \left( \begin{array} {cc}
     1 & {\bf 0} \\  {\bf t} & {\rmT} \end{array} \right)$
where $\rmT$ is a $3 \times 3$ matrix (and ${\bf 0}$ and ${\bf t}$
are row and column vectors respectively) so that
\begin{eqnarray} \label{eq:Trep.equiv}
  \Phi(w_0I + \bw \cdot \sigma) =
   w_0I + ({\bf t} + {\rm T} \bw) \cdot \sigma .
\end{eqnarray}
The ${\bf T}$-matrix corresponding to $\wh{\Phi}$ is ${\bf T}^{\dg}$.

When $\Phi$ is also positivity-preserving, it maps the subspace of
self-adjoint matrices in $\mathcal{M}_2$ into itself, which implies that
$\bf T$ is real.
King and Ruskai \cite{KR} showed that any such map
can be reduced,  via changes of basis in ${\bf C^2}$, to the form
\be \label{eq:Trep}
  {\bf T} = \pmatrix{1& 0& 0& 0 \cr t_1& \lambda_1& 0& 0 \cr
    t_2 & 0& \lambda_2& 0 \cr t_3 & 0& 0& \lambda_3} .
\ee
Because a unitary change of basis
$\rho \raw V \rho V^{\dg}$
on ${\bf C^2}$ is equivalent to a 3-dimensional rotation
${\bf w} \raw R{\bf w}$ acting on the Pauli matrices,
this is equivalent to 
\be \label{eq:KRpolarS}
\Phi(\rho) = U \Big[{\Phi}_{{\bf t}, {\bf \Lambda}} \big(V \rho V^{\dg} \big)
\Big]U^{\dg} ,
\ee
where $U, V \in U(2)$ and  $\Phi_{{\bf t}, {\bf \Lambda}}$
denotes the map corresponding to (\ref{eq:Trep}).
The reduction to (\ref{eq:Trep}) is obtained by applying a
modification of the singular value decomposition to the  $3
\times 3$ real matrix $\rmT$  corresponding to the restriction
of $\Phi$ to the subspace of matrices with zero trace.  
However, the constraint that the diagonalization is achieved
using rotations rather than arbitrary real orthogonal matrices
forces us to relax the usual requirement that the $\lambda_k$ be
positive (so that one can only say that $|\lambda_k|$ are
singular values of
$\rmT$). 
See Appendix A of \cite{KR} for details and Appendix C for some examples
and discussion of subtle issues regarding the signs of $\lambda_k$.
(This decomposition is not unique; the parameters can
 be permuted and the signs of any {\em two} of the 
$\lambda_k$ can be changed by conjugating with a Pauli matrix.  Only the
sign of the product  $\lambda_1 \lambda_2 \lambda_3$ is fixed.  
The canonical example of a positivity preserving map which is not
completely positive, the transpose, corresponds to
$\lambda_k$ taking the values $+1, -1, +1$.  Hence it may be surprising
that  the product  $\lambda_1 \lambda_2 \lambda_3$ can be negative,
as in Example 2(b) of Section \ref{sect:boundary}, and the sign of
this product has an impact on the allowed values of the 
translation parameters $t_k$.)
\par
We call a stochastic map
$\Phi$ {\em unitary} if $\Phi(\rho) = U \rho U^{\dg}$ and sometimes write
$\Gamma_U(\rho)$. It is easy to check that the Kraus representation of a
unitary stochastic map is (essentially) unique, and (\ref{eq:KRpolarS})
can be rewritten as
$\Phi = \Gamma_U \circ \Phi_{{\bf t}, {\bf \Lambda}} \circ \Gamma_V$.

We are interested here in the (convex) set  $\cs$ of stochastic
maps,  i.e., those $\Phi$ that satisfy the stronger condition of complete
positivity. The crucial point about the reduction (\ref{eq:KRpolarS}) is
that $\Phi$ is completely positive  if and only if
${\Phi}_{{\bf t}, {\bf \Lambda}}$ is.  Thus, the question of
characterizing stochastic maps reduces to studying matrices of the form
(\ref{eq:Trep}) under the assumption that $|\lambda_k| \leq 1$ (which
is necessary for $\Phi$ to be positivity preserving).  Of
course, this reduction is not necessarily unique when the $\lambda_k$'s
are not distinct; this will be discussed further in section
\ref{sect:subclass}.

The image of the Bloch sphere of pure states under a map of the
form (\ref{eq:Trep})  is the ellipsoid
\be \label{eq:ellip}
 \left( \frac{ x_1 - {t}_1}{\lambda_1} \right)^2 +
   \left( \frac{ x_2 - {t}_2}{\lambda_2} \right)^2 +
   \left( \frac{ x_3 - {t}_3}{\lambda_3} \right)^2 = 1
\ee
so that eigenvalues $\lambda_k$ define the length of the axes
of the ellipsoid and the vector ${\bf t}$ its center.
The Bloch sphere picture of images as ellipsoids
is useful because it allows one to determine geometrically
the states that emerge with minimal entropy and, roughly,
the states associated with maximal capacity.
Note that a trace preserving map is positivity-preserving
if and only if it maps the Bloch sphere into the ``Bloch ball", defined
as the Bloch sphere plus its interior.
However, not all ellipsoids contained in the ``Bloch ball" correspond to images
of a stochastic map $\Phi$.    It was shown in \cite{AF,KR}
that the $\lambda_k$'s are limited by the inequalities
$(\lambda_1 \pm \lambda_2)^2   \leq   (1 \pm \lambda_3)^2$.
However, even for most allowable choices
of $\lambda_k$, complete positivity restricts (often rather severely)
the extent to which translation of the ellipsoid is possible.
Moreover, characterizing the allowable ellipsoids is not equivalent
to characterizing all stochastic maps because (\ref{eq:ellip})  
depends only on $|\lambda_k|$ while the actual conditions restrict
the choice of signs of $\lambda_k$ as well.
It is worth emphasizing that complete positivity is an extremely
strong condition. In fact whether the map is stochastic or not depends on
the position and orientation of that ellipsoid inside the Bloch sphere
and there are many ellipsoids within the Bloch sphere that do {\em not}
correspond to a completely positive map.

For bistochastic maps, the extreme points are known \cite{AF,KR}
to consist of the maps that conjugate by a unitary matrix and,
in the $\lambda_k$ representation, correspond to four corners
of a tetrahedron.  The maps on the edges of the tetrahedron correspond
to ellipsoids that  have exactly two points in common with 
the boundary of the Bloch sphere.   These maps play a special role and
it is useful to consider them as {\em quasi-extreme points}. We
sometimes call the closure of the set of extreme points the set
of the {\em generalized extreme points} and we then refer to
those  points that are generalized, but not true, extreme points 
as quasi-extreme points.   We will see that for
non-unital maps the extreme points correspond to those maps for
which the translation allows the ellipsoid to touch the
boundary of the sphere at two points (provided one interprets a
single point as a pair of degenerate ones in certain special
cases.)   This is discussed in more detail in section
\ref{sect:boundary}.

\bigskip

%\pagebreak

\subsection{Summary of Results} \label{sect:summ}

We now summarize our results for maps of the form
(\ref{eq:Trep}).    For such maps, it is easy to verify
that a necessary condition for $\Phi$ to be
 positivity-preserving is that
$|t_k|+|\lambda_k| \leq 1$ for $k = 1, 2, 3$.
\begin{thm} \label{thm:cpcond}
A map  $\Phi$  given by {\rm (\ref{eq:Trep})} for which
  $|t_3|+|\lambda_3| \leq 1$ is completely positive if and only if
the equation
\begin{eqnarray}\label{eq:Rphi.cond}
\lefteqn{\pmatrix{ t_1+i t_2 & \lambda_1+\lambda_2
\cr \lambda_1-\lambda_2 & t_1+i t_2}} \\
& = &  \pmatrix{ (1+t_3+\lambda_3)^\frac{1}{2} & 0 \cr 0 &
(1+t_3-\lambda_3)^\frac {1}{2}} \, R_{\Phi} \, 
\pmatrix{ (1-t_3- \lambda_3)^\frac{1}{2} & 0 \cr 0 &
(1-t_3+\lambda_3)^\frac {1}{2}}   \nonumber
\end{eqnarray}
has a solution $R_{\Phi}$ that is a contraction.
\end{thm}

\noindent{\bf Remark:}
When $|t_k|+|\lambda_k| < 1$ (\ref{eq:Rphi.cond}) has the unique solution
\begin{eqnarray}\label{eq:cont.mtrx}
R_{\Phi} = \pmatrix{ \ds{\frac{t_1+it_2}{(1+t_3+\lambda_3)^{1/2}
(1-t_3-\lambda_3)^{1/2}}} &
\ds{\frac{\lambda_1+\lambda_2}{(1+t_3+\lambda_3)^{1/2}
       (1-t_3+\lambda_3)^{1/2}}}
\cr ~~& ~~ \cr  ~~& ~~ \cr
\ds{\frac{\lambda_1-\lambda_2} {(1+t_3-\lambda_3)^{1/2}
(1-t_3-\lambda_3)^{1/2}}} &
 \ds{\frac{t_1+it_2}{(1+t_3-\lambda_3)^{1/2}
(1-t_3+\lambda_3)^{1/2}}} } .
\end{eqnarray}
When $|t_3|+|\lambda_3| = 1$ no solution to (\ref{eq:Rphi.cond}) exists
unless $t_1 = t_2 = 0$ and either $\lambda_1 =  \lambda_2$ or
$\lambda_1 = - \lambda_2$.
 In either case,  it would suffice to let
$R_{\Phi} = \frac{1}{2 \sqrt{|\lambda_3|}}
\pmatrix{ 0  & \lambda_1+\lambda_2  \cr \lambda_1 - \lambda_2 & 0 }$.
However, this matrix is singular. Since the solution to
(\ref{eq:Rphi.cond}) is not unique  in this case, it
 will be more convenient to define $R_{\Phi} $ by the non-singular
matrix
\be \label{eq:eql.mtrx}
R_{\Phi} =  \frac{\lambda_1}{\sqrt{|\lambda_3|}}
   \pmatrix{ 0 & 1 \cr 1 & 0} .
\ee
If $\lambda_3 = 0$ and
$|t_3|+|\lambda_3| = 1$, then $|t_3| = 1$ in which case
$\Phi$ will not be completely positive unless
$t_1 = t_2 = \lambda_1 =  \lambda_2 = 0$; then
it is consistent and sufficient to interpret
$\frac{\lambda_1}{\sqrt{|\lambda_3|}} = 1$
so that $R_{\Phi} = \sigma_x $.

\medskip

We now return to the case $|t_k|+|\lambda_k| < 1$ and
analyze the requirement that $R_{\Phi} $ given by
(\ref{eq:cont.mtrx}) is a contraction.
The requirement that the diagonal elements of $R_{\Phi}^{\dg} R_{\Phi}$ and
$R_{\Phi} R_{\Phi}^{\dg} $ must be $ \leq 1$ then implies that
\be
   (\lambda_1 + \lambda_2)^2 & \leq & (1 + \lambda_3)^2  - t_3^2
   - (t_1^2 + t_2^2) \left(
    \frac{1 + \lambda_3 \pm t_3}{1 - \lambda_3 \pm t_3} \right)
 \label{eq: diag.cond.plus}  \\
 &  \leq & (1 + \lambda_3)^2  - t_3^2  \nonumber \\
  (\lambda_1 - \lambda_2)^2 & \leq & (1 - \lambda_3)^2  - t_3^2
   - (t_1^2 + t_2^2) \left(
    \frac{1 - \lambda_3 \pm t_3}{1 + \lambda_3 \pm t_3} \right)
  \label{eq: diag.cond.neg}  \\
 &  \leq & (1 - \lambda_3)^2  - t_3^2 .\nonumber
\ee
This implies that the Algoet-Fujiwara condition \cite{AF}
\be \label{eq:AFcond}
(\lambda_1 \pm \lambda_2)^2 & \leq & (1 \pm \lambda_3)^2  - t_3^2
\ee
always holds.
This was originally obtained \cite{AF} as a necessary condition
for complete
positivity under the {\em assumption} that $\Phi$ has the form
(\ref{eq:Trep}) with the additional constraint $t_1 = t_2 = 0$.
Moreover, it follows that a {\em necessary} condition for complete
positivity of a map of the form (\ref{eq:Trep}) is
\be \label{eq:AFcond.T3}
(\lambda_1 \pm \lambda_2)^2 & \leq & (1 \pm \lambda_3)^2 .
\ee
Although not obvious from this analysis, (\ref{eq:AFcond.T3}) holds
if and only if it is valid
for any permutation of the parameters $ \lambda_k$.

In addition to a condition on the diagonal, the requirement
$R_{\Phi}^{\dg} R_{\Phi} \leq I$ also implies that 
$\det \big(I - R_{\Phi}^{\dg} R_{\Phi} \big) \geq 0.$  This
leads to the condition
\be \label{eq: det.cond}
 \lefteqn{  \biggl[1-(\lambda_1
^2+\lambda_2^2+\lambda_3^2)-(t_1^2+t_2^2+t_3^2)\biggr]^2 } ~~~~~ \\ & \geq &
4\ \biggl[
\lambda_1^2(t_1^2+\lambda_2^2) +\lambda_2^2
    (t_2^2 + \lambda_3^2) +
\lambda_3^2(t_3^2+\lambda_1^2)-2\lambda_1\lambda_2\lambda_3 \biggr]
\nonumber
\ee

Conditions on the diagonal elements and determinant of
$I - R_{\Phi}^{\dg} R_{\Phi}$ suffice
to insure  $R_{\Phi}$ is a contraction.
Moreover, a direct analysis of the case
$|t_k|+|\lambda_k| = 1$ allows us to extend these inequalities to yield
the following result.
\begin{cor} \label{thm:cpcond.ineq}
A map  $\Phi$  given by {\rm (\ref{eq:Trep})} for which
  $|t_3|+|\lambda_3| \leq 1$ is completely positive if and only if
{\em(\ref{eq: diag.cond.plus})}, {\em(\ref{eq: diag.cond.neg})} and
{\em(\ref{eq: det.cond})} hold,
where {\em(\ref{eq: diag.cond.plus})}  and {\em(\ref{eq: diag.cond.neg})}
are interpreted so that $t_1 = t_2 = 0$ when
$|t_3|+|\lambda_3| = 1$.
\end{cor}

We now discuss those maps $\Phi$ given by
(\ref{eq:Trep}) that are extreme points of $\cs$.

\begin{thm} \label{thm:ext.unit}
Let $\Phi$ be a stochastic map induced by a
matrix  ${\bf T}$ of the form {\rm (\ref{eq:Trep})}.
Then $\Phi$ belongs to the closure of the set of extreme
points of $\cs$ if and only if the matrix $R_{\Phi}$
[as given by {\rm (\ref{eq:cont.mtrx})} or {\rm (\ref{eq:eql.mtrx})}
 above] is  unitary.
\end{thm}

If $\Phi \in \cs$, then it is unitarily
equivalent, in the sense of (\ref{eq:KRpolarS}), to a map of the form
(\ref{eq:Trep}). Moreover, as will be shown in section \ref{sect:change},
that ``equivalence" preserves the property of being a   generalized extreme
point of $\cs$.
Thus, in particular, a necessary condition for
$\Phi$ to be an extreme point of
$\cs$ is that it is  equivalent [in the sense of (\ref{eq:KRpolarS})]
to a map of the form (\ref{eq:Trep}) for which the corresponding
$R_{\Phi}$ is unitary.

Since a cyclic permutation of  indices corresponds to a rotation in
${\bf R}^3$ which can be incorporated into the maps $U$ and $V$ in
(\ref{eq:KRpolarS}),  stating conditions (for membership of $\cs$,  or for
any form of extremality)  in a form in which,  say,
$\lambda_3$ and $t_3$  play a special role does not lead to any real loss of
generality. (Non-cyclic permutations are more subtle; e.g,  a rotation of
$\pi/2$ around the $y$-axis is equivalent to the interchange
$t_1 \leftrightarrow -t_3, \lambda_1 \leftrightarrow \lambda_3$.
However, the sign change on $t_3$ does not affect the conditions
(\ref{eq: diag.cond.plus})-(\ref{eq: det.cond}) above nor
(\ref{eq:AFcond.eql})-(\ref{eq:ext.b}) below which involve either
$t_3^2$  or both signs $\pm t_3$.)
Thus, $\Phi$ is in the closure of the extreme points of $\cs$ if
and only if, up to a permutation of indices,  equality holds in
(\ref{eq: diag.cond.plus}), (\ref{eq: diag.cond.neg}) and (\ref{eq: det.cond}).
This can be summarized and reformulated in the following useful form.

\begin{thm} \label{thm:extpts}
A map  $\Phi$  belongs to the closure of the set of extreme points of $\cs$
if and only if it can be reduced to the form {\rm (\ref{eq:Trep})}
 so that at most one
$t_k$ is non-zero and (with the convention this is $t_3$)
\be \label{eq:AFcond.eql}
  (\lambda_1 \pm \lambda_2)^2 = (1 \pm \lambda_3)^2  - t_3^2 .
\ee
Moreover,  $\Phi$ is extreme unless $t_3 = 0$ and $|\lambda_3| < 1$.
\end{thm}

Adding and subtracting the conditions
(\ref{eq:AFcond.eql}) yields the equivalent conditions
\be
   \lambda_3 & = & \lambda_1 \lambda_2  \label{eq:ext.a} \\
   t_3^2 & = & (1 - \lambda_1^2)(1 - \lambda_2^2) \label{eq:ext.b}
\ee
This leads to a useful trigonometric  parameterization of the
extreme points and their Kraus operators, as given below.

We conclude this summary by noting that we have obtained three equivalent
sets of conditions on the closure of the set of extreme points of $\cs$
after appropriate reduction  to the ``diagonal'' form {\rm (\ref{eq:Trep})}.
\begin{enumerate}
 \renewcommand{\labelenumi}{\theenumi}
    \renewcommand{\theenumi}{\alph{enumi})}
\item Equality in (\ref{eq: diag.cond.plus}), (\ref{eq: diag.cond.neg})
   and (\ref{eq: det.cond}),
\item $t_1 = t_2 = 0$ and (\ref{eq:AFcond.eql}) with both signs,
\item $t_1 = t_2 = 0$, (\ref{eq:ext.a}) and (\ref{eq:ext.b}),
\end{enumerate}
where it is implicitly understood that the the last two conditions
can be generalized to other permutations of the indices.
However, it should be noted that the requirement  $t_1 = t_2 = 0$
is equivalent to requiring that $R_{\Phi}$ is skew diagonal.
This is a strong constraint that necessarily excludes many
unitary $2 \times 2$ matrices.   That such a constrained subclass
of unitary matrices can eventually yield all extreme points is a
result of the fact that we have already made a reduction
 to the special form (\ref{eq:Trep}).

It may be worth
noting that when one of the $\lambda_k = 0$, then (\ref{eq:ext.a})
implies that at least two $\lambda_k$ are zero.  The resulting
degeneracy  allows extreme points that do not necessarily have
the form described in Theorem \ref{thm:extpts}.   Although the 
degeneracy in  $\lambda_k$ permits two $t_k$ to be non-zero, 
  it also permits a reduction to the form of the theorem.
  This is discussed in Section \ref{sect:subclass}
as Case IC.

In section \ref{sect:subclass} we consider the implications
of conditions (\ref{eq:ext.a}) and (\ref{eq:ext.b}) in detail.
For now, we emphasize that the
interesting new class of extreme points are those that also
satisfy
$1 > |\lambda_1| > |\lambda_2| > |\lambda_3|  >   0$.

\bigskip

%\pagebreak

\subsection{Trigonometric parameterization}

The reformulation of the conditions of Theorem \ref{thm:extpts}
as (\ref{eq:ext.a}) and (\ref{eq:ext.b}) implies that when
$\Phi$ is in the closure of the extreme points, the  matrix in
(\ref{eq:Trep}) can be written in the useful form
\be \label{eq:Ttrig}
{\bf T} = \left( \begin{array} {cccc}
 1 & 0 & 0 & 0 \\  0 & \cos u & 0 & 0 \\
   0 &  0 & \cos v & 0 \\
\sin u \sin v &  0 & 0 & \cos u \cos v
\end{array} \right)
\ee
with $u \in [0,2\pi), v \in [0,\pi)$. Extending the range of
$u$ from $[0,\pi)$ to $[0,2\pi)$ insures that the above
parameterization yields both positive and negative values of
$t_3$ as allowed by (\ref{eq:ext.b}).  It is straightforward
to verify that in this
parameterization $\Phi$  can be realized with the
Kraus operators
\be \label{eq:Kraus.trig}
A_{+}  &
 = & \big[\cos \half v \, \cos \half u\big] \, I \, +
\, \big[\sin \half v \, \sin \half u\big] \,  \sigma_z
\nonumber \\  ~~& ~~ &     \\ \nonumber
A_{-} &  = & \big[\sin \half v \, \cos \half u \big] \, \sigma_x  \, -  \,
i   \,\big[ \cos \half v \, \sin \half u \big]\,  \sigma_y \, .
\ee
The rationale behind the subscripts $\pm$ should be clear when we
compute the products $A_{\pm} A_{\pm}^{\dg}$ in Section \ref{sect:proof}.

\bigskip

A similar parameterization was obtained by Niu and Griffiths \cite{NG}
in their work on two-bit copying devices.     One generally regards noise
as the result of failure to adequately control interactions
with  the environment.  However, even classically, noise can also arise
as the result of deliberate ``jamming'' or as the inadvertent
result of eavesdropping as in, e.g., wire-tapping.    One
advantage to  quantum communication is that protocols involving
non-orthogonal  signals provide protection against undetected
eavesdropping.   Any attempt to intercept and duplicate the
signal results leads to errors that may then appear as noise to the
receiver.   The work of Niu and Griffiths \cite{NG}
suggests that the extreme points of the form (\ref{eq:Ttrig}) can be
regarded as arising from an eavesdropper trying to simultaneously
optimize information intercepted and minimize detectable effects
in an otherwise noiseless channel.

This parameterization was also obtained independently
by Rieffel and Zalka \cite{RZ} who, like Niu and Griffiths, considered
maps that arise via interactions between a pair of qubits, i.e.,
maps defined by linear extension of
\be \label{eq:lind}
\Phi(E_1) = T_2 U (E_1 \ot E_2) U^{\dg}
\ee
where $T_2$ denotes the partial trace, $U$ is a unitary matrix
in $\calm_4$ and $E_j$ denotes a projection onto a pure state
in ${\bf C}_2$.  Moreover, Niu and Griffiths's construction showed
that taking the partial trace $T_1$ rather than $T_2$ is equivalent
to switching $\sin$ and $\cos$ in   (\ref{eq:Ttrig}).

Since one can obtain all extreme points of $\cs$ by a reduction
via partial trace on ${\bf C}^2 \ot {\bf C}^2$ one might
conjecture, as Lloyd \cite{Lloyd} did, that any map in $\cs$ can be
represented in the form (\ref{eq:lind}) if $E_2$ is replaced
with  a mixed state $\rho_2$.  However, Terhal, et al \cite{T} showed that
this is false and another counter-example was obtained in \cite{RZ}.
This does not contradict Lindblad's representation in \cite{Lind}
because his construction requires that the second Hilbert space
have dimension equal to the number of Kraus operators.  Thus, a map
that requires four Kraus operators is only guaranteed to have a
representation  of the form (\ref{eq:lind}) on ${\bf C}^2 \ot {\bf C}^4$.

\bigskip

%\pagebreak

\section{Proofs} \label{sect:proof}

\subsection{Choi's results and related work}

Our results will be based on the following
fundamental result of Choi \cite{Choi1,P}.
\begin{thm}\label{thm:choi}
A linear map $\Omega: \mathcal{M}_n \rightarrow \mathcal{M}_n$ is completely
positive if and only if $\Omega \otimes I_{\mathcal{M}_n}$
 is positivity-preserving on $\mathcal{M}_n \otimes
\mathcal{M}_n$ or, equivalently, if and only if
the matrix
\begin{eqnarray}\label{Choi.cond}
       \pmatrix{ \Omega(E_{11}) & \dots &  \Omega(E_{1n})
           \cr \dots & \dots &\dots
           \cr \dots & \dots &\dots
           \cr \Omega(E_{n1}) & \dots &  \Omega(E_{nn})  }
\end{eqnarray}
is positive semi-definite where
$(E_{j,k})_{j,k=1}^n$ are the standard matrix units for $\mathcal{M}_n$
so that {\rm (\ref{Choi.cond})} is in
$ \mathcal{M}_n(\mathcal{M}_n)=\mathcal{M}_{n^2}$
\end{thm}
We are interested in $n = 2$, in which case this condition is equivalent
\cite{H2} to
\be \label{eq:Bellform}
  \big( \Omega \otimes I \big) (B_0) \geq 0
\ee
where $B_0$ is the pure state density matrix that projects onto the
maximally entangled Bell state $\psi_{0}$ which physicists usually write
as
$\psi_{0} = \frac{1}{\sqrt{2}} \,
       \left( \, |00 \ket + |11 \ket \, \right) $, i.e.,
$B_0 =  |\psi_{0} \kb \psi_{0}|$.
It will be useful to define
$\beta(\Omega) \equiv 2 \big( \Omega \otimes I \big)  (B_0) $  to be the
matrix in (\ref{Choi.cond}) and write
\begin{eqnarray}\label{phi1}
\beta(\Omega) = \pmatrix{ A &   C
       \cr C^{\dg} & B  },
\end{eqnarray}
so that $A= \Omega(E_{11})$,  $B= \Omega(E_{22})$
 and $C = \Omega(E_{12}) $.
We now assume that $\Phi$ is trace-preserving, so that its adjoint
$\wh{\Phi}$ is unital, $\wh{\Phi}(I) = I$, and write
\begin{eqnarray} \label{phi2}
 \beta( \wh{\Phi})  =
 \pmatrix{ \wh{\Phi}(E_{11}) &   \wh{\Phi}(E_{12})  \cr
\wh{\Phi}(E_{21}) & \wh{\Phi}(E_{22}) } =
  \pmatrix{ A &   C  \cr C^{\dg} & I-A  },
\end{eqnarray}
where we have exploited the fact that $E_{11} + E_{22} = I$
so that $B = \wh{\Phi}(E_{22}) = \wh{\Phi}(I - E_{11}) = I - A$
when $\Omega = \wh{\Phi}$.  Thus, $0 \leq A\leq I$.

Note that
$ \beta( \Phi) = 0 \iff  \Phi(E_{jk}) = 0 ~ \forall ~ j,k.$ Hence
$\beta$ defines an affine isomorphism between $\cs$ and the image
$\beta(\cs) \subset \mathcal{M}_4$.  In particular,
 there is a one-to-one correspondence between extreme points
of $\cs$ and those of the image $\beta(\cs)$ or $\beta(\wh{\cs})$
where   $\wh{\cs} = \{\wh{\Phi} : \Phi \in \cs \} $  denotes the set of
completely positive maps that are unital.

It is also worth  noting that the matrix
\begin{eqnarray} \label{phistar}
\beta(\wh{\Phi}) = 2\big(\wh{\Phi} \otimes I \big) (B_0) = \pmatrix{
\wh{\Phi}(E_{11}) & \wh{ \Phi}(E_{12})
       \cr \wh{\Phi}(E_{21}) & \wh{\Phi}(E_{22})  }
\end{eqnarray}
is obtained from
$\beta(\Phi)$ by conjugating with a
permutation matrix that exchanges the second and the third coordinate in
${\bf C}^4$ and then taking the complex conjugate, i.e.,
\be \label{eq:perm}
%  \big( \Phi \otimes I \big) (B_0)
\beta(\wh{\Phi})=
\overline{U^{\dagger}_{23} \, \beta(\Phi)
%\big[ \big(\wh{\Phi} \otimes I \big) (B_0) \big]
\, U_{23} }
\ee
where
\bee
U_{23} = U^{\dagger}_{23} = \pmatrix{ 1 & 0 & 0 & 0 \cr
  0 & 0 & 1 & 0 \cr 0 & 1 & 0 & 0 \cr 0 & 0 & 0 & 1}.
\eee
In particular, (\ref{eq:perm}) shows that $\beta(\wh{\Phi})$
is  positive semi-definite if and only if  $ \beta( \Phi)$ is.

\bigskip
For maps of the form (\ref{eq:Trep}), one easily finds
\be \label{eq:beta}
\beta(\Phi) =
\frac{1}{2} \ \pmatrix{
    1+ t_3+\lambda_3 & t_1- it_2 & 0 & \lambda_1 +\lambda_2
\cr t_1+it_2 & 1- t_3-\lambda_3 &   \lambda_1 -\lambda_2 & 0
\cr 0&\lambda_1-\lambda_2 &  1+ t_3- \lambda_3
&t_1- it_2
\cr \lambda_1+\lambda_2 & 0 &t_1+it_2 &1-t_3+\lambda_3} 
\ee
or, equivalently,
\begin{eqnarray}\label{eq:beta.hat}
\beta(\wh{\Phi}) =
\frac{1}{2} \ \pmatrix{
    1+ t_3+\lambda_3 & 0 & t_1 + it_2 & \lambda_1 +\lambda_2
\cr 0 & 1+ t_3-\lambda_3 &   \lambda_1 -\lambda_2 & t_1 +it_2
\cr t_1-it_2&\lambda_1-\lambda_2 &  1- t_3- \lambda_3
&0
\cr \lambda_1+\lambda_2 & t_1- it_2 &0 &1-t_3+\lambda_3}.
\end{eqnarray}
The reason why we concentrate on matrices $\beta(\wh{\Phi})$ (rather than
$\beta(\Phi)$) is that for maps of the form
(\ref{eq:Trep}) the matrices $A$ and $B=I-A$
are diagonal,  which greatly simplifies the analysis.

\bigskip

Our next goal is to characterize elements of $\beta(\wh{\cs})$
and the set of extreme points of  $\beta(\wh{\cs})$.   We will do
this by applying Choi's condition to (\ref{eq:beta.hat}) for which we
will need the following  result.
\begin{lemma}\label{prop:extreme}
A matrix
$M= \pmatrix{ A &   C
       \cr C^{\dg} & B  }
$ is positive semi-definite if and only if $A\ge 0, B\ge 0 $ and
$C=A^\frac{1}{2} \ R \ B^\frac{1}{2}$
for some contraction $R$.
Moreover,  the set of
positive semi-definite matrices with fixed $A$ and $B$ is a convex
set whose extreme points satisfy
$C=A^\frac{1}{2} \ U \ B^\frac{1}{2},$
where $U$ is unitary.
\end{lemma}
The first part of the lemma is well-known,
see, e.g,  \cite{HJ2}, Lemma 3.5.12.
The second part follows easily by using
well-known facts that the extreme points of
the set of contractions in $\mathcal{M}_n$ are unitary
(see, e.g. \cite{Ped} Lemma 1.4.7, \cite{HJ2} Section 3.1, problem 27
or Section 3.2, problem 4 )
and that the
 extreme points of the image of this (compact convex)
set under the affine map
$R \raw A^\frac{1}{2} \, R \, B^\frac{1}{2}$ are
images of extreme points.

When the matrix $M=\beta(\wh{\Phi})$ has the form (\ref{eq:beta.hat}),
then $B = I - A$ and the map $R$ coincides with
$R_{\Phi}$ defined by (\ref{eq:Rphi.cond}) (or by (\ref{eq:cont.mtrx}) if $0 < A <
I$). Accordingly, Lemma
\ref{prop:extreme} will imply that a necessary condition that
$\Phi$ be extreme is that $R_{\Phi}$ is unitary.
In fact,  we have the following equivalence  from which
Theorem \ref{thm:ext.unit} follows.
\begin{thm} \label{thm:ext.unitary}
A map $\Phi$ is a  generalized extreme point of $\cs$ if and only if
the corresponding matrix $M = \beta({\wh{\Phi}})$ defined via {\rm
(\ref{phi2})} is of the form
\be \label{eq:unitary}
M = \pmatrix{A & \sqrt{A} \, U \sqrt{I-A}
    \cr \sqrt{I-A} \, U^{\dg} \sqrt{A} & I-A}
\ee
with $0 \le A \le I$ and $U$
unitary.
\end{thm}
Since the matrices $\{ E_{jk} \}$ form a basis for $\calm_2$, any
positive semi-definite matrix of the form (\ref{phi2}) defines a
completely positive unital map $\wh{\Phi}$ and, hence, a stochastic map
$\Phi$. Thus, it remains only to verify the ``if" part,  i.e., that $U$ 
being unitary implies that  $\Phi$ is a generalized extreme point.

To that end, we shall need some additional
results.   First, we observe that Lemma  \ref{prop:extreme}
implies that $M$ is positive semi-definite if and only if it
can be written in the form (all blocks are $2 \times 2$)
\be \label{eq:block2}
   M = \pmatrix{ A &   C \cr C^{\dg} & B  } =
\pmatrix{\sqrt{A}&  0 \cr 0 & \sqrt{B} }
    \pmatrix{I &  R \cr R^{\dagger}  & I }
      \pmatrix{\sqrt{A}&  0 \cr 0 & \sqrt{B} }
\ee
with $R$ a contraction.  Furthermore, it is well-known and
easily verified that a matrix
$R$ in $\calm_2$ is unitary if and only if
   $E_R := \pmatrix{I &  R \cr R^{\dagger}  & I }$ has rank two.
When $A$ and $B$ are both non-singular, it  follows immediately
that a positive semi-definite matrix $M$ can be written in the form
(\ref{eq:block2}) with $R$ unitary if and only if $M$ has rank two.
In the  singular case some care must be used, but we still have
\begin{lemma} \label{lemm:rkM}
Let $M \in \calm_4$  with $M \ge 0$.  Then $M$ admits a factorization
{\rm (\ref{eq:block2})} with $R$ unitary if and only
if
${\rm rank}(M) \leq 2$.
\end{lemma}
\noindent{\bf Proof:}  The argument is quite standard (e.g., it can be
extracted from results in \cite{HJ2}, in fact for blocks of any size)
but we include it for completeness. We need
to show that if ${\rm rank}(M) \leq 2$,  then the equation
\be \label{eq:c}
   C = \sqrt{A} \, R \, \sqrt{B}
\ee
admits a unitary solution $R$. This holds, by the comments above,
if $A$ and $B$ are nonsingular and,  trivially,  if $A=0$ or $B=0$.
To settle the remaining cases we observe that by conjugating $M$ with
a matrix of the form  $\pmatrix{ V & 0 \cr 0 & W}$ where  $V, W$
are $2 \times 2$ unitaries,  we may reduce the question to the
case when $A$ and $B$  are diagonal. If  ${\rm rank}(A)={\rm rank}(B)=1$,
the equation (\ref{eq:c}) imposes restriction on only one of the entries
of $R$.  By the first part of Lemma \ref{prop:extreme} that entry must be
$\le 1$ in absolute value for $M$ to be positive-definite,  and the
remaining entries can be  chosen so that $R$ is unitary. If,  say,
${\rm rank}(A)=2$ and ${\rm rank}(B)=1$ (say, only the first diagonal
entry of $B$ is $\neq 0$), then ignoring the last row and last column of $M$
(which must be  0)  we can think of equation (\ref{eq:c}) as involving
nonsingular matrices $A \in \calm_2$ and $B \in \calm_1$ and a $2 \times 1$
matrix $R$. The condition that ${\rm rank}(E_R) \leq 2$ then implies 
that $R$ is a norm one column vector  (if $R$ was an $n \times m$ matrix
with $m \le n$,  the condition for ${\rm rank}(E_R) \leq n$ would be
that $R$ is an isometry).  Returning to $2 \times 2$ blocks we notice that in
the present case the equation (\ref{eq:c}) does not impose any restrictions on
the {\em second } column of $R$ and so that column can be chosen so
that $R$ is unitary.  (See, e.g., the
discussion of unitary dilations on pp. 57-58 of \cite{HJ2}.) \qed

\bigskip We will also need more
results from Choi \cite{Choi1}. First,  there is a special case of
Theorem 5 in \cite{Choi1} which we state in a form adapted to our
situation and notation.
\begin{lemma}  \label{lemma:choi2}
A stochastic map   is an extreme
point of $\cs$ if and only if it can be written in the form
{\em (\ref{eq:kraus})} (necessarily  with $\sum_k A_{k}  A_{k}^{\dg} =
I$), so that the set of matrices $\{A_{j} A_{k}^{\dg} \}$ is linearly
independent.
\end{lemma}
In addition to being an important ingredient in the proof of Theorem
\ref{thm:ext.unitary}, this result will allow us to distinguish between
true extreme points and those in the closure.

\bigskip The next result is ``essentially" implicit in \cite{Choi1}.
\begin{lemma}  \label{lemma:choi3}
The minimal number of Kraus operators $A_k$ needed to represent
a completely positive map in the form
{\rm (\ref{eq:kraus})} is ${\rm rank}[\beta(\Phi)]$.
\end{lemma}
Indeed,  Choi showed that  one
can obtain a set of Kraus operators
for any completely positive map $\Phi$ from an orthogonal
set of eigenvectors of
$\beta(\Phi)$ corresponding to non-zero eigenvalues.
Hence one can always write $\Phi$
in the form (\ref{eq:kraus}) using only  rank$[\beta(\Phi)]$ Kraus
operators.  The other direction is even simpler.
It is readily checked (and it also follows from
the proof of Choi's Theorem \ref{thm:choi}) that if $\Phi(\rho) := V
^{\dg} \rho V$ with $V \neq 0$, then ${\rm rank}\, \beta(\Phi) = 1$
and so,  in general,
${\rm rank}\, \beta(\Phi)$ does not exceed the number of Kraus operators.

The above results together with yet unproved Theorems \ref{thm:ext.unitary}
and \ref{thm:extpts}
allow us to compile a list of conditions characterizing the closure of
extreme points of the set $\cs$ of stochastic maps on $\calm_2$ (the
``generalized extreme points").   These conditions are given in
Theorem
\ref{thm:meta} of section \ref{sect:change}.

\medskip

%\pagebreak

\subsection{Proof of Results in Section \ref{sect:summ} }

\noindent{\bf Proof of Theorem \ref{thm:cpcond} and  Remark:}
By Theorem \ref{thm:choi}, it suffices to show that (\ref{eq:beta.hat})
is positive semi-definite.  To do this we
apply Lemma \ref{prop:extreme} to (\ref{eq:beta.hat}).
Since the corresponding $A$ and $B$ are then diagonal, they are positive
semi-definite if and only if each term on the diagonal is non-negative
which is equivalent to $|t_3| + |\lambda_3| \leq 1.$
If this inequality is strict, then both $A$ and $B$ are
invertible and it follows easily that
$R = A^{-1/2} C B^{-1/2}$ has the form given by (\ref{eq:cont.mtrx}).

If either or both of $A,B$ is singular,  we use the fact that when a
diagonal element of an $n \times n$ matrix is zero, the matrix
can be positive semi-definite only if the corresponding
row and column are identically zero.  Thus, if any of the
diagonal elements of (\ref{eq:beta.hat}) is zero, then
$t_1 = t_2 = 0$ and one of $\lambda_1 \pm  \lambda_2 = 0$.
It is then straightforward to check that
$R_{\Phi}$  can be chosen to have the form in  (\ref{eq:eql.mtrx}).
The cases $\lambda_1 =0 $ and $  \lambda_3 = 0$ can also be
easily checked explicitly, which suffices to verify  the remark
following Theorem \ref{thm:cpcond}.
\qed

\medskip

The next proof will use Lemma \ref{lemma:choi2} which requires
the following matrix products, which are easily computed from
(\ref{eq:Kraus.trig}).
\be
2A_{+} A_{+}^{\dg} & = &
 \big[1 + \cos u \cos v \big]
\, I   +  \big[
  \sin u \sin v \big] \, \sigma_z
  \nonumber \\
2A_{-} A_{-}^{\dg} & = &
\big[1 - \cos u \cos v \big]
\, I   -  \big[
  \sin u \sin v \big]  \, \sigma_z \nonumber  \\
2A_{\pm} A_{\mp}^{\dg} & = &
   \big[ \sin  v \big] \, \sigma_x \, \pm i \,
       \big[ \sin u \big] \, \sigma_y.
\label{eq:kraus.prod}
\ee
Note that it follows immediately that
$A_{+} A_{+}^{\dg} + A_{-} A_{-}^{\dg}= I$ as required for a
trace-preserving map $\Phi$.

We also point out that when a completely positive map $\Phi$ can be
realized with one Kraus operator $A$, the condition
$\sum_k A_{k}  A_{k}^{\dg} = I$ reduces to $A A^{\dg} = I$, from which
it is elementary that $A^{\dg} A = I$ as well.   Thus, when a map $\Phi$
can be realized with a single Kraus operator $A$, then $\Phi$ is
unital $\iff$ $\Phi$ is trace-preserving $\iff$  $A$ is unitary.

\medskip

\noindent{\bf Proof of Theorems \ref{thm:ext.unit} and \ref{thm:extpts}:}
We argue as above and observe that fixing $A, B$ in Lemma  \ref{prop:extreme}
when applied to $M=\beta(\wh{\Phi})$ is equivalent to fixing
$\wh{\Phi}(\sigma_z)$ or the last row of ${\bf T}$ (note also that
$A$ and $B$ are diagonal if and only if the last row of ${\bf T}$ is as in
(\ref{eq:Trep}), i.e., if its two middle entries are 0).  It then follows
immediately from the second part of Lemma
\ref{prop:extreme} that when $\Phi$ is an extreme point
$R_{\Phi}$ must be unitary.  By continuity and compactness, this holds also
for  $\Phi$'s in the closure of the extreme points.

We now  show that this unitarity implies condition (\ref{eq:AFcond.eql})
in Theorem \ref{thm:extpts}.  First, we observe that when $R_{\Phi}$ is
unitary, equality holds in both (\ref{eq: diag.cond.plus})
and (\ref{eq: diag.cond.neg}).  When $t_3 \neq 0$ and $\lambda_3 \neq 0$, this
is possible only if $t_1 = t_2 = 0$, which yields (\ref{eq:AFcond.eql}).

When $| t_3| + |\lambda_3 |  =1$ (including the possibilities
$t_3 = 0,   |\lambda_3 |  =1$ and $t_3 = 1,   |\lambda_3 |  =0$)
the necessity of the conditions
in Theorem \ref{thm:extpts} can be checked explicitly using
the observations in the previous proof.

 When $t_3 = 0$ and $0 < |\lambda_3 | < 1$,
the necessity of the conditions in Theorem \ref{thm:extpts}
requires more work.  The condition that the columns of $R_{\Phi}$
are orthogonal becomes
\bee
  (t_1 + i t_2) \frac{\lambda_1 - \lambda_2}{1 - \lambda_3 } +
    (t_1 - i t_2) \frac{\lambda_1 + \lambda_2}{1 + \lambda_3 } = 0
\eee
which implies
\bee
  t_1 \, \left[ \frac{\lambda_1 - \lambda_2}{1 - \lambda_3 } +
    \frac{\lambda_1 + \lambda_2}{1 + \lambda_3 } \right] & = & 0 \\  ~~ \\
 t_2 \, \left[ \frac{\lambda_1 - \lambda_2}{1 - \lambda_3 } -
    \frac{\lambda_1 + \lambda_2}{1 + \lambda_3 } \right] & = & 0 .
\eee
Both quantities in square brackets can not simultaneously be zero
unless $\lambda_1 = \lambda_2 = 0$.   Although we can then have
$t_1, t_2 \neq 0$ this is, except for permutation of indices an
allowed special case of (\ref{eq:AFcond.eql}), as discussed in
Section \ref{sect:subclass} as case (IC).   If only one of the
quantities in square brackets is zero, then we again have only one
$t_k$ non-zero, and (\ref{eq:AFcond.eql}) holds after a suitable
permutation of indices.  Suppose, for example, that $t_2 \neq 0$.
Then
\be \label{eq:ext.a2}
\frac{\lambda_1 - \lambda_2}{1 - \lambda_3 } =
    \frac{\lambda_1 + \lambda_2}{1 + \lambda_3 }
  ~ \imp ~ \lambda_2 = \lambda_1 \lambda_3
\ee
and
\be \label{eq:ext.b2}
 \lefteqn{ 1 = \frac{t_2^2}{ 1 - \lambda_3^2} +
 \frac{(\lambda_1 + \lambda_2)^2}{(1 + \lambda_3)^2 }
  = \frac{t_2^2}{ 1 - \lambda_3^2} +
  \frac{\lambda_1 + \lambda_2}{1 + \lambda_3 }
 \frac{\lambda_1 - \lambda_2}{1 - \lambda_3 }}
   ~~~~~~~~~~~~~~~~~~~~   \nonumber \\
& \imp &  \nonumber
   1 - \lambda_3^2 = t_2^2 + \lambda_1^2 - \lambda_2^2
  = t_2^2 + \lambda_1^2 -  \lambda_1^2 \lambda_3^2 \\
 & \imp & (1 - \lambda_3^2)(1 -\lambda_1^2) = t_2^2 .
\ee
Except for interchange of $t_2, \lambda_2 $ with $t_3, \lambda_3$,
equations
 (\ref{eq:ext.a2}) and (\ref{eq:ext.b2}) are equivalent
to (\ref{eq:ext.a}) and (\ref{eq:ext.b}).

The case $\lambda_3 = 0$ can be treated using an argument similar to
that above for $t_3 = 0$.  Thus, we have verified that unitarity
of $R_{\Phi}$ implies
(\ref{eq:AFcond.eql}) in all cases.

Conversely, one can easily check  that under the hypotheses
$t_1 = t_2 = 0$ and  (\ref{eq:AFcond.eql}),
 $R_{\Phi}$ as given by
(\ref{eq:cont.mtrx}) or (\ref{eq:eql.mtrx}) is always unitary.
The remaining question is whether or not every unitary matrix
$R_{\Phi}$ gives rise to a generalized extreme point of $\cs$.
To answer this question we use the
fact that the corresponding maps  can be written in the
form (\ref{eq:Ttrig}) with Kraus
operators given by  (\ref{eq:Kraus.trig}).
We then make  the following series of observations.

\begin{itemize}

\item[I)] Because the set $\{ I, \sigma_x, \sigma_y, \sigma_z \}$
 forms an orthonormal basis (with respect to the Hilbert-Schmidt inner
  product)
for $\mathcal{M}_2$, the operators in (\ref{eq:kraus.prod}) are
linearly independent if and only if  {$\sin  v \,  \sin  u \neq 0$}.
Thus, by Lemma \ref{lemma:choi2}, the corresponding
stochastic map $\Phi$ is extreme if {\em both} of
$\{\sin  v, \,  \sin  u \}$ are non-zero
or, equivalently, $t_3 \neq 0$ and all $|\lambda_k| < 1$.

\item[II)]
If  {\em both} of
$\{\sin  v, \,  \sin  u \}$
are zero, then
$u,  v$ are both of the form $n \pi$ for some integer $n$.
In this case one and only one of the two Kraus operators given by
(\ref{eq:Kraus.trig}) is non-zero and, hence, unitary.
In this case, $\Phi$ is unitary and obviously extreme, and
$\lambda_k = \pm 1$ for  $k = 1,2,3$.  In fact, the non-zero
Kraus operator is simply one of  $\{ I, \sigma_x, \sigma_y, \sigma_z \}$
and, accordingly, either none or exactly two of the
$\lambda_k$ can be negative.

\item[III)] If exactly one of
$\{\sin  v, \,  \sin  u \}$ is zero, then
it follows from (\ref{eq:Ttrig}) that
$t_3 = \sin v  \sin u = 0$, and $\Phi$ is unital.
The extreme points of the unital maps are known \cite{AF,KR2},
and given by (II) above.  Hence  maps for which exactly one of
$\{\sin  v, \,  \sin  u \}$ is zero can not be
extreme.  However,  it is readily verified that such a $\Phi$ is
a convex combination of two extreme maps of type (II).

\end{itemize}
Since these three situations exhaust the possible situations
in (\ref{eq:Ttrig}), they also cover all possible
unitary maps which arise from an $R_{\Phi}$ of the form under consideration.
 Moreover, maps of type (III) form a 1-dimensional
subset of the 2-dimensional torus given by the
parameterization (\ref{eq:Ttrig}) and so they must be in the closure of the
set of extreme points. Maps of type (I) and (II) correspond
to the non-unital and unital extreme points of $\cs$ respectively.
\qed

\medskip
The argument presented thus far yields
the extreme points of $\cs$ that are represented by a matrix
{\bf T}  of the form (\ref{eq:Trep}).  Since every $\Phi \in \cs$ is
equivalent in the sense of (\ref{eq:KRpolarS}) (i.e.,  after changes of
bases in the 2-dimensional Hilbert spaces corresponding to the domain and
the range of $\Phi$) to a map  of the form (\ref{eq:Trep}),  this describes
{\em in principle } all extreme points of
$\cs$  and all points in the closure of the set of extreme  points.

Although it is possible to characterize the extreme points of $\cs$
without appealing to the special form (\ref{eq:Trep}), we chose the
present approach for two reasons. One is that the use of the diagonal form
considerably simplifies the computations.  The other is that it leads to a
parameterization that is useful in applications.  Nevertheless, this
basis dependent approach may obscure some subtle issues that we now
point out.

If $R_{\Phi}$ is unitary, but does not have the special
form (\ref{eq:cont.mtrx}), it will still define a stochastic map,
 albeit {\em not} one of the ``diagonal'' form (\ref{eq:Trep}).
However, even though Theorem \ref{thm:ext.unitary} implies that the
corresponding stochastic map $\Phi$ is a generalized extreme
point of $\cs$, we are not able to deduce that immediately.
The difficulty is that we do not know that the property of the matrix
$R$ [implicitly defined by (\ref{eq:block2})] of being unitary is
invariant under changes of bases,  and in particular whether it yields an
unitary $R_{\Phi}$ after reduction to the form (\ref{eq:cont.mtrx}).   In
the next subsection, we clarify that issue and complete
the proofs of Theorem \ref{thm:ext.unitary} and Theorem \ref{thm:meta}
that follows.

\bigskip

\subsection{Invariance of conditions under change of basis }
\label{sect:change}

We begin by carefully describing the result of a change of basis
on $\beta(\Phi)$ or $\beta(\wh{\Phi})$.   It will be convenient
to let $\Gamma_A$ denote conjugation with the  matrix $A$,
i.e., $\Gamma_A(\rho) = A \rho A^{\dg}$.  Then for any pair of
unitary matrices $U, V$,
$\Phi(\rho) = U \big[{\Phi}_D \big(V \rho V^{\dg} \big) \big]U^{\dg}$
is equivalent to $\Phi = \Gamma_U \circ \Phi_D \circ \Gamma_V$
or   $\Phi_D = \Gamma_{U^{\dg}}\circ \Phi\circ \Gamma_{V^{\dg}}$.
Notice that the map 
$
\Phi= \Gamma_U \circ \Phi_D \circ \Gamma_V$ is an affine isomorphism of $\cs$ and
in particular it preserves the sets of extreme and quasi-extreme points.
We will be primarily interested in the case  when this reduces
a map $\Phi$ to diagonal form $\Phi_D$  as in (\ref{eq:KRpolarS}).
However, our results apply to any pair
of positive maps related by such a change of basis.
\begin{lemma} \label{lemma:change}
Let $\Phi$ be a positive map on $\calm_2$ and $U, V$  unitary.
Then $\Phi = \Gamma_U \circ \Phi_D \circ \Gamma_V$ if and only if
\be \label{eq:basis}
  (\Phi \ot I) (B_0) =
   [U \ot V^T] \, [(\Phi_D \ot I) (B_0)] \, [U^{\dg} \ot \overline{V}]
\ee
where where $V^T$ denotes the transpose and
{\em [}as in {\em (\ref{eq:Bellform})]} $2B_0$
is the matrix with blocks $E_{jk}$.
\end{lemma}
\noindent{\bf Proof:} First observe that,
for any set of matrices $\{ G_{jk} \}$ in $\calm_2$,
\bee
  \big[(\Gamma_U \circ \Phi_D) \ot I \big]
   \pmatrix{ G_{11} & G_{12} \cr G_{21} & G_{22}} =
\pmatrix{ U & 0 \cr 0 & U}
   \pmatrix{ \Phi_D(G_{11}) & \Phi_D(G_{12})
          \cr \Phi_D(G_{21}) & \Phi_D(G_{22}) }
               \pmatrix{ U^{\dg} & 0 \cr 0 & U^{\dg}}.
\eee
The result then follows from the  fact that
\be \label{eq:ent.trans}
(\Gamma_V \ot I) (B_0) = (I \ot \Gamma_{V^T}) (B_0)
\ee
which can be verified directly in a number of ways. \qed

It may be worth remarking that (\ref{eq:ent.trans}) extends to the
general case in which $B_0$ is replaced by the $n^2 \times n^2$
matrix with blocks $E_{jk}$ and may characterize maximally
entangled states.

\medskip

We now present the promised
list of conditions characterizing the closure of
extreme points of the set $\cs$ of stochastic maps on $\calm_2$.

\begin{thm} \label{thm:meta}
For a  map $\Phi \in \cs$ the following conditions are equivalent:
\begin{itemize}
\item[{\rm (i)}] $\Phi$ belongs to the closure of the set of
extreme points of $\cs$.

\item[{\rm (ii)}]  $M = \beta({\wh{\Phi}})$ can be written in
the form {\rm (\ref{eq:unitary})} {\rm [or, equivalently, in the form
(\ref{eq:block2})]} with $R$ unitary and $B=I-A$.

\item[{\rm (iii)}]\  $\Phi$ can be reduced via changes of bases
{\rm (\ref{eq:KRpolarS})} to a map of the form {\rm (\ref{eq:Ttrig})}.

\item[{\rm (iv)}]\  $\Phi$ can
be represented in the form  {\rm (\ref{eq:kraus})} using not more than two
Kraus operators.

\item[{\rm (v)}] \ ${\rm rank}\, \beta({\wh{\Phi}}) =  {\rm
rank}\, \beta(\Phi) \le 2$.
\end{itemize}
\end{thm}
\noindent {\bf Proof.} We start by carefully reviewing the arguments
presented thus far.  In the preceding section we showed that,  in the
special case when
$\Phi$ is represented by a matrix {\bf T}  of the form (\ref{eq:Trep}),
the first three conditions are equivalent.  Moreover,  the implication
(i) $\Rightarrow$ (ii) holds in full generality.  Concerning the other
conditions {\em in the general case}, Lemma \ref{lemm:rkM} says that (ii)
$\Leftrightarrow {\rm rank}\,
\beta({\wh{\Phi}}) \le 2$, while it follows from (\ref{eq:perm}) that
${\rm rank}\, \beta({\wh{\Phi}}) =  {\rm
rank}\, \beta(\Phi)$; thus (ii) $\Leftrightarrow$ (v).  Similarly,
Lemma  \ref{lemma:choi3} shows that (iv) $\Leftrightarrow$ (v).
In addition, Lemma \ref{lemma:change} above implies that (v) is
invariant under a change of basis.  It
thus remains to prove that (ii) $\Rightarrow$ (iii) $\Rightarrow$ (i) in
full generality rather than just for maps of the form (\ref{eq:Trep}).
But this follows from the facts that every stochastic map can be reduced
to the form (\ref{eq:Trep}) via  changes of bases  (\ref{eq:KRpolarS}) and
that all these conditions are invariant under such changes of bases.  The
latter was already noticed for (i) and is trivial for (iii).  To show the
(not obvious at the first sight) invariance for (ii) we notice that
we have already proved that (ii) $\Leftrightarrow$ (iv) and that (iv)
is clearly invariant under changes of bases:  if
$\Phi_1(\rho) = \sum_k A_{k}^{\dg} \rho A_{k}$ and
$\Phi(\rho) = U \Big[{\Phi_1} \big(V \rho V^{\dg} \big) \Big]U^{\dg}$,
then $\Phi(\rho) = \sum_k \big(V^{\dg}A_{k} U^{\dg}\big)^{\dg} \rho
\big(V^{\dg}A_{k} U^{\dg}\big) $.  \qed

\medskip \noindent {\bf Remark}. It is possible
to directly characterize the extreme points (rather than generalized extreme
points)  if the map is not of the form (\ref{eq:Trep}). However, the
conditions thus obtained are not very transparent.

\medskip

Since Theorem \ref{thm:ext.unitary} is ``essentially"  (i)
$\iff$ (ii) above, we have also proved that result.
 The following result gives another approach to proving Theorem
\ref{thm:ext.unitary} since it easily implies (i) $\iff$ (iv).
Although redundant, we include it because its proof (which should be
compared to the argument in the preceding section) is of independent
interest.
 \begin{thm}  \label{thm:two.kraus}
A stochastic map $\Phi$ that, when written in the form (\ref{eq:kraus}),
requires two Kraus operators $A_k$ but not more, is either an extreme
point of $\cs$ or bistochastic.
\end{thm}
\noindent{\bf Proof:}  Since $\Phi$ trace-preserving implies
$\sum_k A_{k}  A_{k}^{\dg} = I$, we can assume without loss of
generality that $A_1  A_1^{\dg} = D$ and $A_2  A_2^{\dg} = I-D$
where $D$ is diagonal and $0 < D < I$.  Then we can write
$A_1 = \sqrt{D} \, V_1,~ A_2 = \sqrt{I-D}\, V_2$ with $V_1, V_2$ unitary.
By Lemma \ref{lemma:choi2}, $\Phi$ is extreme if and only if the set
$\{A_1 A_1^{\dg}, \,  A_2 A_2^{\dg}, \, A_1 A_2^{\dg}, \, A_2 A_1^{\dg} \}$
is linearly independent.  This is equivalent to linear independence of
the set
$$ \Big\{ \, D, ~ I-D, ~ \sqrt{D} \,
     W \sqrt{I-D},~ \sqrt{I-D} \, W^{\dg} \sqrt{D}\, \Big\} . $$
where $W = V_1 V_2^{\dg}$ One can readily verify that this set is
linearly independent unless $D$ is a multiple of the identity or
$W$ is diagonal.  If $D = \mu I$, then both $\mu^{-1/2} A_1$
and $(1-\mu)^{-1/2} A_2$ are unitary so that
$\Phi$  is a convex combination of  unitary maps.  The second
exception is more subtle.  We first note that the fact that $W$
is diagonal implies
$ A_1^{\dg} A_1 = V_2^{\dg} W^{\dg} D W V_2 = V_2^{\dg} D V_2$.
Thus
\bee
  A_1^{\dg} A_1 + A_2^{\dg} A_2 = V_2^{\dg} D V_2 + V_2^{\dg} (I-D) V_2 = I
\eee
so that $\Phi$ is a bistochastic map.   \qed

\medskip

One might think that a map of the form (\ref{eq:Trep})  with all three
$t_k$  non-zero would require three extreme points.  However,
this is not the case.   In general, two maps in the set of
generalized points suffice.  If a bistochastic map is written
in diagonal form, then it
has a unique decomposition as a convex combination of the four 
unitary maps corresponding to the corners of a tetrahedron; however, it
can also be written non-uniquely as a convex combination of two maps on
the ``edges'' of the tetrahedron.   For non-unital maps, not only do two
extreme points suffice, but they can be chosen so that an arbitrary
non-unital map is the ``midpoint'' of a line connecting two (true)
extreme points.
\begin{thm} \label{thm:2maps}
Any stochastic map on $\mathcal{M}_2$ can be written as the 
convex combination of
 of two maps in the closure of the set of extreme points.
\end{thm}
This is an immediate consequence of  Theorem \ref{thm:ext.unitary}
and the following elementary result.   Note that in both cases we
 have proven the somewhat stronger result that the convex combination
can be chosen to be a midpoint.
\begin{lemma}
Any contraction in $\mathcal{M}_2$ can be written as the 
convex combination of two unitary matrices.
\end{lemma}
\pf If $R$ is a contraction, its singular value decomposition
can be written in the form
\be \label{eq:conv.decomp}
R & = &  V \pmatrix{ \cos \theta_1 & 0 \cr 0 & \cos \theta_2} W^{\dg}
 \nonumber  \\ ~ \nonumber \\ & = &
   \half V \pmatrix{ e^{i \theta_1} & 0 \cr 0 & e^{i \theta_2}} W^{\dg} +
  \half V \pmatrix{ e^{-i\theta_1} & 0 \cr 0 & e^{-i\theta_2}} W^{\dg}
\ee
where $V$ and $W$ are unitary.   \qed

\medskip

Note that if $R = V D W^{\dg}$ as above, then
\bee
\lefteqn{
\pmatrix{ A & \sqrt{A} R \sqrt{B} \cr \sqrt{B} R^{\dg} \sqrt{A} & B }  }
\\ & = & \pmatrix{ V &  0 \cr 0 & W }
   \pmatrix{ V^{\dg} A V &   \sqrt{V^{\dg} A V} D \sqrt{W^{\dg} B W }
   \cr \sqrt{W^{\dg} B W} D^{\dg} \sqrt{V^{\dg} A V}   &  W^{\dg} B  W }
   \pmatrix{ V^{\dg} &  0 \cr 0 & W^{\dg} }
\eee
However, this transformation does {\em not} correspond to a change of
basis of the type considered at the start of this section.

\bigskip

%\pagebreak

\section{Discussion and Examples} \label{sect:discuss}

\subsection{Types of extreme points} \label{sect:subclass}

We begin our discussion of extreme points by using the
trigonometric parameterization (\ref{eq:Ttrig}) to find
image points that lie on the Bloch sphere.   We consider
a pure state of the form $\rho = \half[I + \bw \dtsig]$ with
$\bw = ( \pm \cos \theta, 0, \sin \theta)$ so that
$\Phi(\rho) = \half[I + \bx \dtsig]$ with
$\bx = ( \pm \cos \theta  \cos u, 0, \sin u \sin v + \sin \theta
  \cos u \cos v)$.  After some straightforward trigonometry, we find
\be \label{eq:bound}
|\bx|^2 & = & \cos^2 \theta  \cos^2 u +  \sin^2 \theta  \cos^2 u \cos^2 v
  + \sin^2 u \sin^2 v  \nonumber \\
& ~ & ~~~~~~~+ 2 \sin \theta \, [\sin u \sin v  \cos u \cos v]  \nonumber \\
  & = & 1 - [\sin u \cos v - \sin \theta \cos u \sin v]^2
\ee
so that $|\bx| = 1$ if (and only if)
$\sin \theta = \frac{\tan u }{ \tan v}$.  This will be possible
if $|\tan u| \leq |\tan v|$ or, equivalently,
$|\lambda_1| \geq |\lambda_2|$.  In particular,  $\Phi$
maps the pair of states corresponding to
$( \pm \cos \theta, 0, \sin \theta)$ on the Bloch sphere
to the pair of states
corresponding to $( \pm \cos \omega, 0, \sin \omega)$
when
\be \label{eq:trig.map}
 \cos \omega & = & \cos \theta \,  \cos u = \frac{
        \sqrt{ \cos^2 u \sin^2  v -  \sin^2  u \cos^2 v } }{ \sin v }\\
  \sin \omega & = & \sin u \sin v + \sin \theta \cos u \cos v
      = \frac{\sin u }{ \sin v }.
   \nonumber
\ee

We now describe several subclasses in the closure of the extreme points
described in Theorem \ref{thm:extpts}, using the categories defined
in  its proof.  We make the additional
assumption that $|\lambda_1| \geq |\lambda_2|$ since, in any case,
all permutations of indices must be considered to obtain the
extreme points of all maps of the form(\ref{eq:Trep}).
With that understanding, the list below is exhaustive.
A general extreme point is then the composition of a map of
type (I) or (II)  with unitary maps, as in (\ref{eq:KRpolarS}).

\begin{itemize}

\item[I)]
$1 > |\lambda_1| \geq |\lambda_2| > |\lambda_3| > 0$,
   $t_1 = t_2 = 0$ and
$t_3^2 = (1 - \lambda_1^2)(1 - \lambda_2^2)$.  This class includes
all non-unital extreme points, and can be subdivided to
distinguish some important special cases.
\begin{itemize}

\item[A)] $ |\lambda_1| > |\lambda_2| > 0$ can be regarded as
  the generic situation.   The image $\Phi(\cd)$
   is an ellipsoid translated orthogonal to its major axes
  until exactly two points touch the Bloch sphere, as described above
  and shown in Figure 1.

\item[B)]  $|\lambda_1| = |\lambda_2| > 0$.
  As $|\lambda_1| \raw|\lambda_2|$ the two image points on the
Bloch sphere merge so that  when
$\lambda_1 =  \pm \lambda_2 $,  we recover the amplitude-damping channel
with one fixed point at the  North or South pole
corresponding to $t_3 = \pm (1 - |\lambda_3|)$.

\item[C)] $\lambda_2 = 0 \imp \lambda_3 = 0$.
  If $\lambda_1 \neq 0$, the image $\Phi(\cd)$ is a
line segment whose endpoints lie on the Bloch sphere.  Moreover, the
degeneracy $\lambda_2 = \lambda_3 = 0$ permits a rotation
so that $t_2 \neq 0$ provided that
$t_2^2+t_3^2 = 1-\lambda_1^2$.

If $\lambda_1=0$ as well, we have a completely
noisy channel with all $\lambda_k = 0$.
The image $\Phi(\cd)$ consists of a single point
on the unit sphere onto which all density matrices are mapped,
and the degeneracy permits
a translation ${\bf t}$ by  vector of length
$|{\bf t}| = 1$ in an arbitrary direction.
\end{itemize}

\item[II)] All $|\lambda_k|=1$, ${\bf t} = 0 $.
  In this case, either zero or two of the $\lambda_k = -1$ and
the others $+1$.  As discussed in Appendix B of \cite{KR},
  these four possibilities correspond to 4 points of a tetrahedron.
 Each of these extreme points is a map
with exactly one Kraus operator corresponding to the identity
or one of the three Pauli matrices.  $\Phi$ takes $\cd$ {\em onto}
itself, i.e., $\Phi(\cd) = \cd$.

\item[III)] $\lambda_1 = \pm 1$, $\lambda_2 = \pm \lambda_3 = \mu$ with
  $|\mu |< 1$.  In this case we must have ${\bf t} = 0$, and
  $\Phi$ is not a true extreme point, but a point on an ``edge''
  formed by taking the convex combination of two corners of
   the tetrahedron (II) above.  The  image $\Phi(\cd)$ is an
ellipsoid whose major axis has length one.  Thus, $\Phi(\cd)$ has one
  pair of orthogonal states on the unit sphere, which are also
  fixed points.

These maps have two non-zero Kraus
operators and, hence,  the form
$s \Phi_j + (1-s) \Phi_k$ where  $j,k \in \{0, 1,2,3\}$
and $\Phi_j (\rho) = \sigma_j \rho \sigma_j$ (and $\sigma_0 = I$).
For each  pair $(j,k)$ we obtain a line between two of the
extreme points that form the tetrahedron of bistochastic
maps.

\end{itemize}

It is  worth pointing out that case (IC) is the
{\em only} situation in which one can have
 extreme points with more than one $t_k$ nonzero.
When $\Phi$ has the form (\ref{eq:Ttrig}) this can only
happen when $\lambda_3$ is degenerate [i.e., equal to
$\lambda_1$ or $\lambda_2$.]  However, this is precluded
by  (\ref{eq:ext.a})  unless $\lambda_3 = 0, \pm1$.  In the
latter case we must have a unital channel with all $t_k = 0$.
Thus,  two nonzero  $t_k$ occur only in channels that are
so noisy that at least two $\lambda_k = 0$.

%\bigskip

%\pagebreak

\subsection{Images of stochastic maps} \label{sect:boundary}

The discussion at the start of Section \ref{sect:subclass} suggests
that, roughly speaking, extreme points correspond to maps for
which two (or more) points in the image
$\Phi(\cd)$ lie on the Bloch sphere as shown in Figure 1.  However, this
statement is correct only if we interpret the single point at a pole
 as a pair of degenerate images when $|\lambda_1| = |\lambda_2|$ .

\begin{figure}
{\hspace{4cm}
\epsfxsize=7cm
\epsfbox{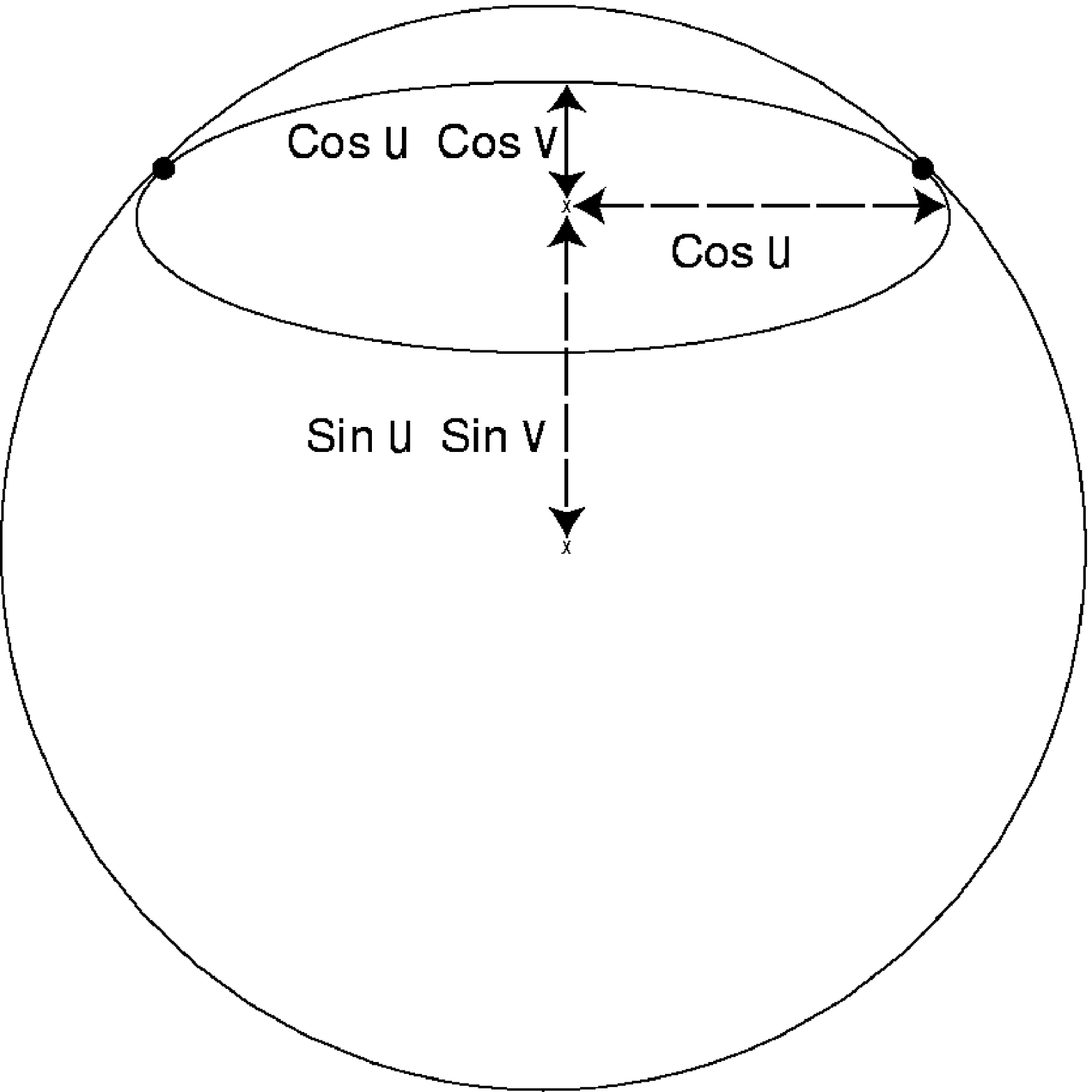}
\caption{Cross section of the Bloch sphere and its image
for an
extreme point of the form (\ref{eq:Ttrig}).  The cross section
is shown in the
plane for which two points on the ellipsoid touch the sphere;
and the axes lengths and shift of the center are indicated under
the assumption that $\cos u > \cos v$.}
}
\end{figure}

Now suppose we want to find the map(s) $\Phi$ that take the
pair of points $( \pm \cos \omega, 0, \sin \omega)$
on the Bloch sphere to the pair $( \pm \cos \theta, 0, \sin \theta)$.
 Geometric considerations and the arguments leading to (\ref{eq:trig.map})
imply that this will be possible only if
$|\sin \theta| > |\sin \omega|$ and  $|\lambda_1| > |\lambda_2|$.
For simplicity, we assume $0 < \omega < \theta < \half \pi$
[corresponding to an upward translation and $u \in (0, \pi)$]
and seek solutions satisfying $0 < u < v < \half \pi$
(corresponding to $\lambda_1 > \lambda_2 > 0$).
The conditions
\bee
  \cos u = \frac{\cos \omega}{\cos \theta},  ~~
     \sin v = \frac{\sin u}{\sin \omega}
\eee
follow from (\ref{eq:trig.map}) and imply that the solution
is unique.

Note that  $\cos \omega > 0, ~ \cos \theta < 0$ will yield solutions
with $\half \pi < v < u < \pi$ which includes  a rotation
of the ellipsoid by $\pi$ as well as a translation, or
$( \pm \cos \omega, 0, \sin \omega)$
to $( \mp |\cos \theta|, 0, \sin \theta)$.

Maps that take pairs of the form $(0,  \pm \cos \theta,  \sin \theta)$
to $(0,  \pm \cos \omega,  \sin \omega)$ require
$|\lambda_2| \geq |\lambda_1|$ and can be treated similarly.
All other situations can be reduced to these after suitable
rotations.  Hence, it follows
that when two distinct extreme points have a common pair of
points on the image of the Bloch sphere,  their pre-images
must be distinct.
\begin{thm}  If $\Phi$ is an extreme point of $\cs$, then at
least one point in the image $\Phi(\cd)$  is a pure state.
If $\Phi(\cd)$ contains two or more pure states, it must
be in the closure of the extreme points of $\cs$.  Moreover, if
the intersection of $\Phi(\cd)$ with the Bloch sphere consists
of exactly two non-orthogonal pure states, then $\Phi$ must be a
non-unital extreme point; while  $\Phi(\cd)$ contains  the
entire Bloch sphere if and only if it is a unital extreme point.
\end{thm}

A non-extreme point can have an image point on the
boundary of the Bloch sphere only if all its extreme components
have the same pre-image for that point. Indeed, if $\Phi = \alpha
\Phi_1 + (1 - \alpha) \Phi_2$ with
$0 < \alpha < 1$ is not extreme and $\Phi$ has an image point
$\bx=\Phi(\bu)$ on  the boundary of the Bloch sphere, then
it follows immediately from
the strict convexity of the Euclidean unit ball that
$\bx=\Phi(\bu)=\Phi_1(\bu)=\Phi_2(\bu)$.

 One can generate a large class of such maps from any pair of extreme
points by applying a suitable ``rotation" to one of them.
Hence, one can find many non-extreme maps that take one
(but not two) pure states to the boundary of the Bloch sphere.
A few special cases are worth particular mention.

\medskip

%\pagebreak

 \noindent {\bf 1. Example of non-extreme maps that reach the boundary:}
\begin{enumerate}
    \renewcommand{\labelenumi}{\theenumi}
    \renewcommand{\theenumi}{\alph{enumi})}

\item  $\lambda_1 = \lambda_2  = \lambda_3 = \mu$
 with $0 \leq \mu < 1$.
This corresponds to a depolarizing channel in which case $\Phi$ maps
the unit ball into a sphere of radius $\mu$ which can then be translated
in {\em any} direction with $|{\bf t}| \leq 1 - |\mu|$.
Similar remarks hold when one $\lambda_k = +\mu$
and the other two $ -\mu$, since conjugating a stochastic map
with one of the Pauli matrices, yields another
 stochastic map with the signs of any {\em two} $\lambda_k$ changed.

\item  Let $|\lambda_1 | = |\lambda_2 | = \mu$,
$ |\lambda_3| \geq \mu^2$, and $|t_3| = (1 \pm \lambda_3)$
so that $R_{\Phi}$ is given by (\ref{eq:eql.mtrx}).
 This is possible if, in addition,
$\lambda_3$ and $\lambda_1 \lambda_2$ have the same sign,
and  $t_1 = t_2 = 0$.  Under these conditions $R_{\Phi}$,
is always a contraction, but is unitary only for $ |\lambda_3| = \mu^2$.
However, the condition $|t_3| = (1 \pm \lambda_3)$ implies
that the North or South pole is a fixed point.  Hence, even when
$ |\lambda_3| > \mu^2$ so that $\Phi$ is not extreme, the
ellipsoid is shifted to the boundary in a manner analogous
to an amplitude damping channel.

\end{enumerate}

In general, however, it is not possible to translate the
image ellipsoid to the boundary of the unit ball.   On the
contrary, there are many situations in which the translation
is severely limited despite contraction of the ellipsoid.

In discussing this question the pair of inequalities (\ref{eq:AFcond})
play an important role and it is natural to ask if
 (\ref{eq:AFcond}) can be extended to
\be  \label{eq:AFcond.gen}
 (\lambda_1 \pm \lambda_2)^2 \leq  (1 \pm \lambda_3)^2  - |{\bf t}|^2
\ee
when more than one $t_k$ is non-zero. By rewriting
(\ref{eq: diag.cond.plus} ) and (\ref{eq: diag.cond.neg} )
in the form
\bee
   (\lambda_1 + \lambda_2)^2 & \leq & (1 + \lambda_3)^2   - |{\bf t}|^2
   - (t_1^2 + t_2^2) \left(
    \frac{2 \lambda_3}{1 - \lambda_3 \pm t_3} \right)  \\
  (\lambda_1 - \lambda_2)^2 & \leq & (1 - \lambda_3)^2  - |{\bf t}|^2
   + (t_1^2 + t_2^2) \left(
    \frac{2 \lambda_3 }{1 + \lambda_3 \pm t_3} \right)
\eee
one can see that, depending on the sign of $\lambda_3$  one of the pair
inequalities in (\ref{eq:AFcond.gen}) holds.   However, in general
we do not expect that (\ref{eq:AFcond.gen}) will hold with both signs.

\medskip

%\pagebreak

 \noindent {\bf 2. Example of maps with limited translation:}
\begin{enumerate}
    \renewcommand{\labelenumi}{\theenumi}
    \renewcommand{\theenumi}{\alph{enumi})}

\item
 If $(\lambda_1 \pm \lambda_2)^2 = (1 \pm \lambda_3)^2$ holds with
exactly one sign, then  $\lambda_1, \lambda_2, \lambda_3$ lie on
the surface of  the tetrahedron defined by (II), but on the
interior of one of the four faces.   However, even when
equality holds with only one sign, (\ref{eq:AFcond}) implies
that ${\bf t} = 0$ so that no translation is possible.  Thus,
one can find maps with  all $|\lambda_k| < 1$ for which the image
$\Phi(\cd)$ is an ellipsoid strictly contained within the unit ball
but no translation in any direction is possible.

\item
A map with $\lambda_1 = \lambda_2 = \lambda_3 = \mu < 0$  is
{\em not}
completely  positive unless  $|\mu| \leq  \thrd$. However, unlike
the case with $\mu > 0$, such maps can
not be translated to the boundary; in fact, for $\mu = - \thrd$, one must
have ${\bf t} = 0$ so that no translation is possible.

\item
When $\lambda_1 = \lambda_2 = \mu, \, \lambda_3 = 0$
which implies $|\mu| \leq \half$ and the image of $\Phi$ is a
circle in a plane parallel to the equator.
Moreover, the equations following (\ref{eq:AFcond.gen})  imply
\be
  4 \mu^2 \leq (1 - |{\bf t}|^2)
\ee
when $\lambda_3 = 0$.
Thus, when $\mu = \half$ no translation the image $\Phi(\cd)$
is possible despite the fact that the ellipsoid has shrunk to
a flat disk of radius $\half$.
For $0 < \mu < \half$ translation is possible, but clearly never
reaches the boundary of pure states.

\end{enumerate}

Example (2a) includes maps with
$\lambda_1  = \lambda_2 = \mu$, $\lambda_3 = 2\mu - 1$,
which may seem to contradict  Example (1b).
However, the condition $2\mu - 1 = \lambda_3 > \lambda_1^2 = \mu^2$
is never satisfied since if would imply
$- \mu^2 + 2\mu - 1 = - (1 - \mu)^2 > 0$.
 Hence the conditions for these two examples do not overlap.

Example (2b) illustrates that the ellipsoid picture, while
useful, is incomplete.  The eigenvalues $[\thrd,\thrd,\thrd]$
and $[\thrd,\thrd,-\thrd]$ both yield the same ellipsoid (actually,
a sphere of radius $\thrd)$. The former can be translated
in any direction with $|{\bf t}| \leq 2/3$; however, the latter
can not be translated at all without losing complete positivity.

\medskip

It is tempting to think of a stochastic map $\Phi$ as
the composition
of a rotation, contraction (to an ellipsoid), translation,
and another rotation.  However, this is not accurate in the
sense that the individual maps in this process will no
longer be stochastic.
In general, it is {\em not} possible to write a non-unital
map as the composition of a unital map (which contracts the
Bloch sphere to an ellipsoid) and a translation.  Such
a translation would  have to be representable in the form
$\pmatrix{ 1 & {\bf 0} \cr \bt & I}$ which does not
satisfy the conditions for complete positivity. 

\bigskip

%\pagebreak

\subsection{Geometry of stochastic maps} \label{sect:geom}

In the previous section we discussed the geometry of the images of
various types of stochastic maps.
We now make some remarks about the geometry of the set of stochastic
maps $\cs$ itself.

\medskip

In a fixed basis, the convex set of stochastic maps with $\bt = 0$
forms a tetrahedron that we can describe using the $\lambda_k$
in $(\ref{eq:Trep})$ which we write as
 $[\lambda_1,\lambda_2,\lambda_3]$.  The extreme points of
the tetrahedron are
$[1,1,1], [1,-1,-1], [-1,1,-1], [-1,-1,1]$.  The edges connecting
any two of these extreme points have the form $[\pm 1, s, \pm s]$
up to permutation and correspond to quasi-extreme points
of type (III).   Although this tetrahedron lies in a 3-dimensional
space in a fixed basis, an arbitrary bistochastic map requires
9 parameters that could be chosen  either as the 9 elements of the real
$3 \times 3$ matrix $\rmT$, or as the three $\lambda_k$ and two
sets of Euler angles corresponding to the two rotations required
to reduce $\Phi$ to diagonal form.  In a fixed basis, an arbitrary
bistochastic map can be written (uniquely) as a convex combination
of 4 true extreme points, or non-uniquely as a convex combination of
two quasi-extreme points on the ``edges''.  As the changes of
basis rotate the tetrahedron (considered as a set of diagonal $3 \times
3$ matrices in ${\bf R}^9$), we obtain a much larger set so that
(by Theorem \ref{thm:2maps}) any
non-extreme  map can be written as the midpoint of two
``generalized'' extreme points  in different bases.

We now describe  the convex set analogous to the tetrahedron
(in a fixed basis) for non-zero ${\bf t}$ with ${\bf t}$ fixed.
We will consider the case when $t_1=t_2=0$ and $t_3 >0$ is fixed  and the
$\lambda_k$ vary. Thus, we are in a 3-dimensional submanifold of
${\bf R}^6$.  It follows from (\ref{eq:Ttrig}) that the
extreme points are given by the curve
$[\cos u, \cos v, \cos u \cos v]$ subject to the constraint
$\sin u \sin v = t_3$.
This can be written in parametric form as a pair of curves
\be \label{ext.curve}
  [\cos u, \pm \cos \left( \sin^{-1}\Big[\frac{t_3}{\sin u} \Big]\right),
 \pm  \cos u  \, \cos \left( \sin^{-1}\Big[\frac{t_3}{\sin u}
\Big]\right)]
\ee
with $\sin^{-1}(t_3) \leq u \leq \pi - \sin^{-1}(t_3)$.  This
curve forms the boundary of Figure 3 below.
Letting $\alpha^2 = 1 - t_3$, we see that the curve (\ref{ext.curve})
passes through the four points
$[\alpha, \alpha, \alpha^2]$,  $[-\alpha, -\alpha, \alpha^2]$,
$[\alpha, -\alpha, -\alpha^2]$, $[-\alpha, \alpha, -\alpha^2]$
when $\sin u = \pm \sin v$.   Thus, as $t_3$ moves away from zero, the
corners of the tetrahedron are replaced by these points
as shown in Figure 3 which might be described as an asymmetric
rounded tetrahedron.   We have
shown  the tetrahedron for comparison, placed in such a way that the
edge connecting the vertices $[-1,-1,1]$ and $[1,1,1]$ is pointing
towards us, and oriented the rounded tetrahedron similarly.
The 4 points above split the curve into four pieces, corresponding
to four of the six edges of the tetrahedron. In place of  the remained
two edges there appear two segments (still on the surface of the
``rounded" tetrahedron)   connecting pairs of the four above points  for
which
$\lambda_3$ has the same sign.   Unlike the case $t_3 = 0$ these
lines are not extreme points, even in a generalized sense.

\begin{figure}
\includegraphics*{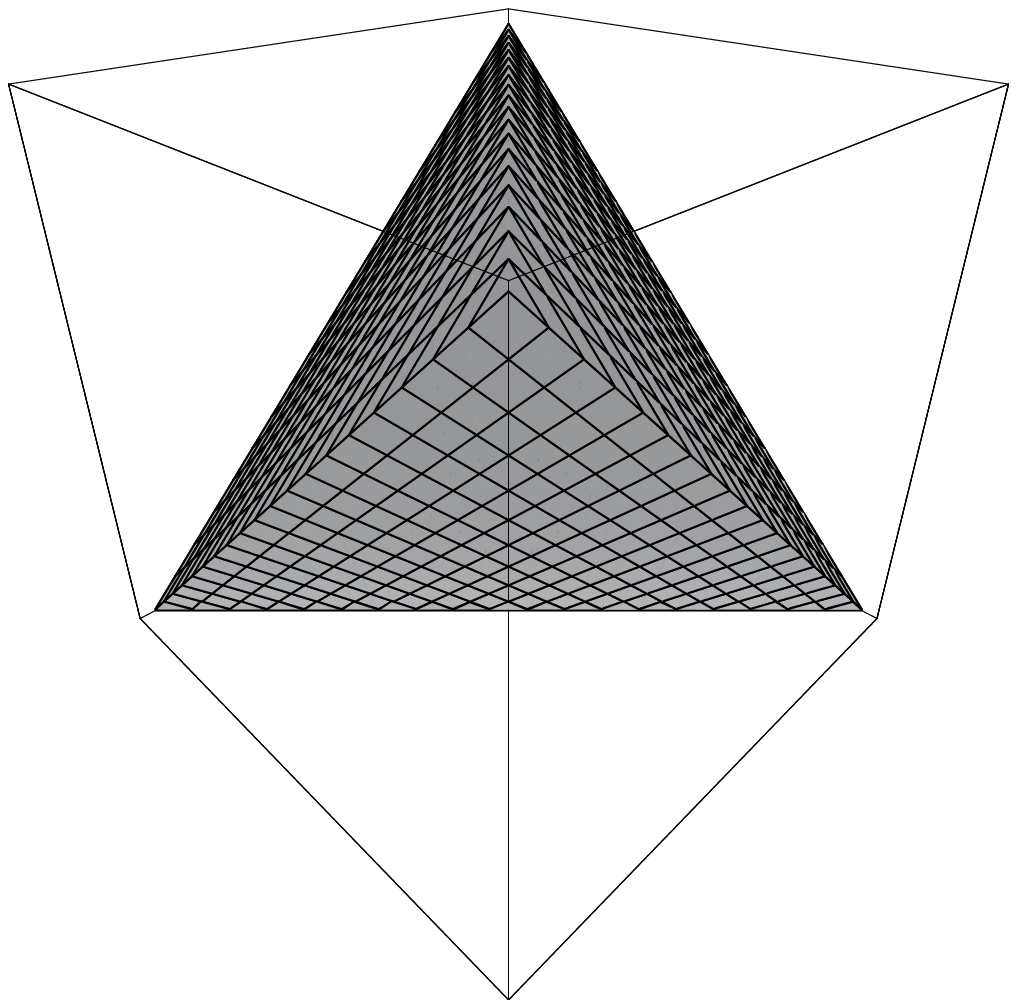}
\caption{the  tetrahedron}
\end{figure}

\begin{figure}
\includegraphics*{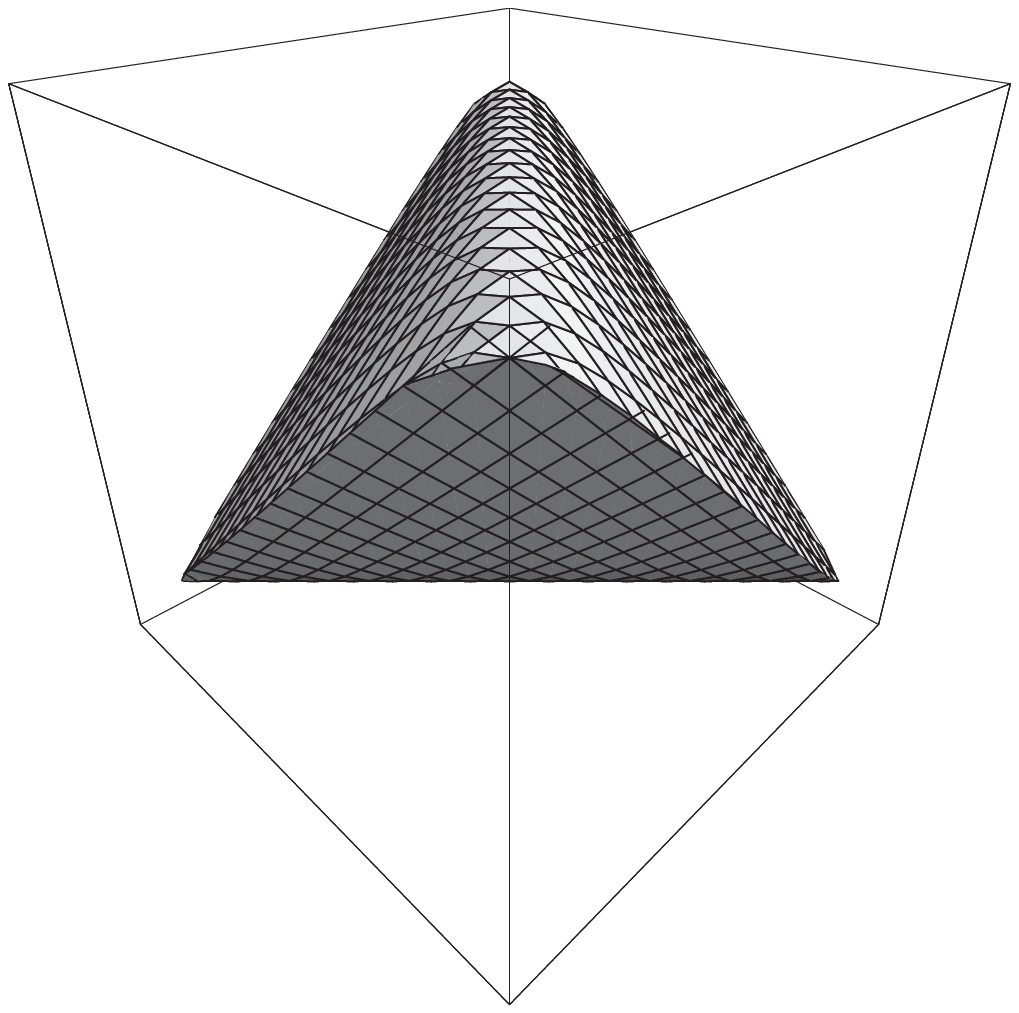}
\caption{the ``rounded tetrahedron"}
\end{figure}

The inequalities (\ref{eq:AFcond}) imply that, as for the tetrahedron,
the figure in (3b) has rectangular cross sections for fixed $\lambda_3$
with   $|\lambda_1 \pm \lambda_2 | \leq \sqrt{(1+\lambda_3)^2 - t_3^2}$.
When $t_3 \neq 0$, the corners depend non-linearly on $\lambda_3$,
yielding a curve of true extreme points.  When $t_3 = 0$, the linearity
in $\lambda_3$ yields a line segment so that we no longer have ``true"
extreme points.

The cases ${\bf t} = (\sqrt{1 - \alpha^2}, 0 , 0)$ and
${\bf t} = (0 , \sqrt{1 - \alpha^2}, 0)$ are similar.
The only difference is the orientation and displacement, e.g.,
whether the point [1,1,1] is replaced by  $[\alpha^2, \alpha, \alpha]$
or to $[\alpha, \alpha^2, \alpha]$.
\bigskip

%In order to see the effect of these relative shifts we plot
%all three extremal curves on one set of axes.   The convex hull of
%these points is a subset of the convex set
%\bee
% \left\{ [\lambda_1, \lambda_2, \lambda_3] :  \Phi_{{\bf t},{\bf
%\Lambda}}
%\in \cs , ~ t_k \geq 0, ~ \hbox{and} ~ \sum_{k=1}^3 t_k =
%%  \sqrt{1 - \alpha^2}   \right\}.
%\eee
%\bigskip

%Fig. 4     All 3 extreme curves plotted together for
%$\alpha = 1, \frac{\sqrt{3}}{2}, \half$

%      If we could find a way to distinguish the curve from Fig. 2 in
%some way, we could eliminate that figure, and one could how Fig. 3b
%is embedded in this one -- further clarifying the main purpose of
%this figure -- showing the relative orientation of the three cases.

%\bigskip
%Each of the curves (\ref{ext.curve}) can be thought of as composed
%of four pieces, each corresponding to one of the edges of the
%original tetrahedron.   This yields a total of twelve curves, with
%each of the original edges corresponding to two distinct curves
%with different $t_k \neq 0$.

\bigskip

%\pagebreak

\subsection{Channel capacity} \label{sect:capac}

The capacity of a quantum communication channel depends on the
particular protocols allowed for transmission and measurement as
well as the noise of the channel.  See \cite{BS1,KR2} for
definitions and  discussion of some of these.  We consider here only the
so-called {\em Holevo capacity } that corresponds to
communication using product signals and entangled measurements
and is now believed (primarily on the basis of numerical
evidence) to be the maximum capacity associated with communication
that does not involve prior entanglements.
The Holevo capacity is given by
\begin{eqnarray}\label{Holv.cap}
  C_{\holv}(\Phi) = \sup_{\pi_j, \, \rho_j} \left(
  S[ \Phi(\rho)] - \sum_j \pi_j  S[ \Phi(\rho_j) ] \right),
\end{eqnarray}
where $S(P) = -\tr \, (P \,\log P)$ denotes the von Neumann entropy
of the density matrix $P$,  $\rho_j$ denotes a
set of pure state density matrices,  $\pi_j$ a discrete
probability vector, and $\rho = \sum_j \pi_j \rho_j$.
It is easy to see that for extreme points of the type
(II) and (III), $C_{\holv}(\Phi) = \log 2$ so that
the capacity attains its maximal value, and for the completely
noisy channel in (IC), $C_{\holv}(\Phi) = 0$.

By contrast, the non-unital situation (I) is more interesting because
it includes channels for which the capacity (\ref{Holv.cap})is strictly
bigger than the classical Shannon capacity, i.e., the
capacity for communication restricted to
product input and measurements.  Such channels demonstrate a definite
quantum advantage. Moreover, the capacity is, in general, achieved
{\em neither} with a pair of orthogonal input states,
which would yield
$ h \big(\sin u \sin v) - h(\sqrt{1 - \sin^2 u \cos^2 v} \big)$,
{\em nor} with
a pair of minimal entropy states, which would yield
$h \big(\frac{\sin u }{\sin v}\big)$
where
\be
 h(t) = - \frac{1 + t}{2} \log \frac{1 + t}{2}
    - \frac{1 - t}{2} \log \frac{1 - t}{2} .
\ee
This behavior was observed first by Fuchs for a non-extreme
channel \cite{Fu}; subsequently, the amplitude-damping channel
was studied by Schumacher and Westmoreland \cite{SW}.
In both cases the work was numerical, but suggests that this
situation is generic for channels that are translated
orthogonal to the major axis of the ellipse.

In the case (IC) when the ellipsoid shrinks to a line, the
capacity can be computed explicitly as
\be  \label{eq:capac.binary}
C_{\holv}(\Phi) & = & h(\sqrt{t_1^2 + t_2^2} =
   h(\cos u ) \\
    & > &  \log 2 - h(\sin u) \nonumber
\ee
since $h(\cos \theta) + h(\sin \theta) \geq \log 2$.
The expression $\log 2 - h(\sin u) $   is the capacity
of the unshifted channel.  In this case, it is also the classical
Shannon capacity.
Holevo \cite{Ho1,Ho2} introduced  such channels
and showed that  they suffice to demonstrate
the ``quantum advantage'' mentioned above, although the both
the minimal entropy and capacity are achieved  with orthogonal
inputs.

\bigskip

\noindent {\bf Remark:}
Holevo  called  channels of the form (IC) ``binary'' because
their Kraus operators can be represented in the form
$ A_1 = |e_1 \kb \psi_+ |,~ A_2 = |e_2 \kb \psi_- |$
where the states $|e_k \ket$ are orthonormal and $\psi_{\pm}$
can, in principle, be any pair of states, such as
corresponding to $( \pm \cos x,   \sin x)$.
When $e_1, e_2$ are  the
eigenvectors $\frac{1}{\sqrt{2}}(1, \pm 1)$ of $\sigma_x$,
the resulting Kraus operators are
$$ A_1 = \frac{1}{\sqrt{2}}\pmatrix{ \cos x & \sin x \cr \cos x & \sin x},
~~  A_2 = \frac{1}{\sqrt{2}}\pmatrix{ -\cos x & \sin x \cr \cos x & -\sin
x}$$ One can easily check that this yields a map of type (IC)
with $u = 2x-\frac{\pi}{2}, v = \half \pi$ but the Kraus operators do not
have the form (\ref{eq:Kraus.trig}).

\bigskip

%\pagebreak

\noindent {\bf Acknowledgment:}   It is a pleasure to thank
Professor Chris King,  Professor Denes Petz and Dr. Barbara Terhal
for stimulating and helpful discussions,  Dr. Christopher
Fuchs for bringing reference \cite{NG} to our attention,
and Professor Roger Horn for clarifying remarks about Lemma
\ref{lemm:rkM}.   After an earlier
version of this paper \cite{RSZ} was posted we learned that
Dr. Eleanor Rieffel and Dr. C. Zalka had independently obtained
some similar results about the extreme points.
We are grateful to them for a number of useful comments,
including drawing our attention to Choi's criterion for
extremality which helped to streamline the proofs, and
raising the question of the minimal number of extreme points
needed for a given map.  MBR would also like to thank Professor
Ruedi Seiler for his hospitality at the Technische Universit\"at
Berlin, where part of this paper was written.
The pictures were done with Mathematica and Adobe Illustrator.
We want to thank Joachim Werner for his help in creating them.

\bigskip

%\pagebreak

\end{document}